# Future Computer Systems and Networking Research in the Netherlands: A Manifesto

ICT Research Platform Netherlands, SIG Future Computer and Network Systems

Editors: Alexandru Iosup (VU, A.Iosup@vu.nl) and Fernando Kuipers (TU Delft, F.A.Kuipers@tudelft.nl)

## Executive summary

**Society's engine for a responsible and sustainable future? Computer Systems!** Our modern society and competitive economy depend on a strong digital foundation and, in turn, on sustained computer systems research and innovation. Computer systems, ranging from small, embedded devices to large data centers and the networks that connect them, are a remarkable technology area with an outstanding impact on society. This manifesto focuses on the Netherlands, where data centers and related ICT infrastructure enable over 3.3 million jobs and over 60% of the GDP, and novel services and products, and where every €1 invested in such systems generates €15 in added value. Sustained investments in capable networking infrastructure have made the Netherlands home to one of the largest data hubs in the world. Cloud adoption exceeds 90% for economic organizations, and 65% for government and public education organizations. Our overarching goal with this document is to highlight the grand societal, technological, and scientific opportunities and challenges in future computer systems and networking (the CompSys area), and to outline how to maintain the leading position the Netherlands has in this area.

**Future-proof digitalization requires ICT research and development (R&D), now.** The Dutch Government and societal stakeholders have identified an urgent need to expand economic and social activities by leveraging and integrating ICT in their knowledge, processes, expertise, and capabilities. This approach, of profound *digitalization*, can make current activities much more valuable, enable new development that can transform society, and lead to unprecedented efficiency and thus universal access to ICT for the first time in history. Without action, the Netherlands could lose its leadership role in the CompSys area, ICT services may lose their flexible, trusted, and secure status in the digital economy, ICT researchers in the Netherlands could be brain-drained to other countries, and the existing shortage of ICT-skilled people could grow. We are not alone: our main allies and competitors in Europe, including Germany and France, and in the world, including the US and China, are intensifying their strategic long-term R&D investments in ICT, with strong emphasis on systems and networking infrastructure. In Europe, understanding is growing that ICT must uphold European values.

**With this manifesto, we draw attention to CompSys as a vital part of ICT.** Among ICT technologies, CompSys covers all the hardware and all the operational software layers that enable applications; only application-specific details, and often only application-specific algorithms, are not part of CompSys. Each of the Top Sectors of the Dutch Economy, each route in the National Research Agenda, and each of the UN Sustainable Development Goals pose challenges that cannot be addressed without groundbreaking CompSys advances. Looking at the 2030-2035 horizon, important new applications will emerge only when enabled by CompSys developments. Triggered by the COVID-19 pandemic, millions moved abruptly online, raising infrastructure scalability and data sovereignty issues; but governments processing social data and responsible social networks still require a paradigm shift in data sovereignty and sharing. AI already requires massive computer systems which can cost millions per training task, but the current technology leaves an unsustainable energy footprint including large carbon emissions. Computational sciences such as bioinformatics, and "Humanities for all" and "citizen data science", cannot become affordable and efficient until computer systems take a generational leap. Similarly, the emerging quantum internet depends for the foreseeable future on (traditional) CompSys to bootstrap operation. Large commercial sectors, including finance and manufacturing, require specialized computing and networking or risk to become uncompetitive. And, at the core of Dutch innovation, promising technology hubs, deltas, ports, and smart cities, could see their promise stagger due to critical dependency on non-European technology.

**Computer systems must be manageable, responsible, sustainable, and usable.** To fulfill the promise of our digitalized future, and avoid the risk of becoming technologically dependent, the Netherlands must address now four grand **societal challenges** related to CompSys:

1. **Manageability**: How to tame the ever-growing complexity, and related human error, in massive computer systems and networks?
2. **Responsibility**: How to realize responsible computer infrastructure – analogous to the concept of responsible AI – whose operation we can rely on (from privacy, ethical, security, performance, availability, and durability perspectives)?
3. **Sustainability**: How to reduce and make the most from the energy footprint of (increasing levels of) computing?
4. **Usability**: How to work, within the context of Europe's digitalization ambitions, with the critical sectors for the Dutch society, such as industry, healthcare, and governance, to make computer systems usable and accessible by all? How to enable AI and quantum computing?

**Addressing these challenges requires an innovative, multi-year vision and approach…**

**… through a groundbreaking, holistic, and collective technology roadmap**
The Netherlands is well-positioned to be a big player in CompSys, a consequence of its rich legacy in communications and networking, and of its current community of CompSys researchers. Because we are at the end of Moore's Law, and because our society has just started deep digitalization, the time is ripe for groundbreaking advances in all layers of computer systems and networking, see Figure 1. Because advances in computer systems are inextricably linked, it is imperative the community takes a *holistic view* of computer systems and networks. These fundamental topics require further, equally fundamental advances in *how* we do science in the field, and demand collective thinking and action.

**… through renewed education, truly accessible for all**
As a major contribution to the Dutch society, we need a concerted action to reintroduce *systems thinking* as a clear curricular goal, helping our society understand and shape the increasingly interconnected and complex digitalized world. We need IT professionals with core computer systems expertise, coming from all backgrounds and from under-represented groups.

**… and through a concerted answer to our call to action**
In the race to digital empowerment, we see a clear opportunity for Europe and the Netherlands to not focus on only one aspect of digitalization, e.g., AI algorithms, but to consider CompSys technology holistically. Motivated by societal needs and encouraged by the strengths of CompSys research in the Netherlands, we call for a large-scale collective effort and investment in this field. The benefits of digitalization are enormous and reward every area of our economy and society. However, if we fail to do this soon, there is a high risk that the CompSys technology and specialists required by our society will not be available in the Netherlands anymore, further limiting our *digital sovereignty*. Economically, *the Netherlands cannot out-compete without computer systems and networking research!*

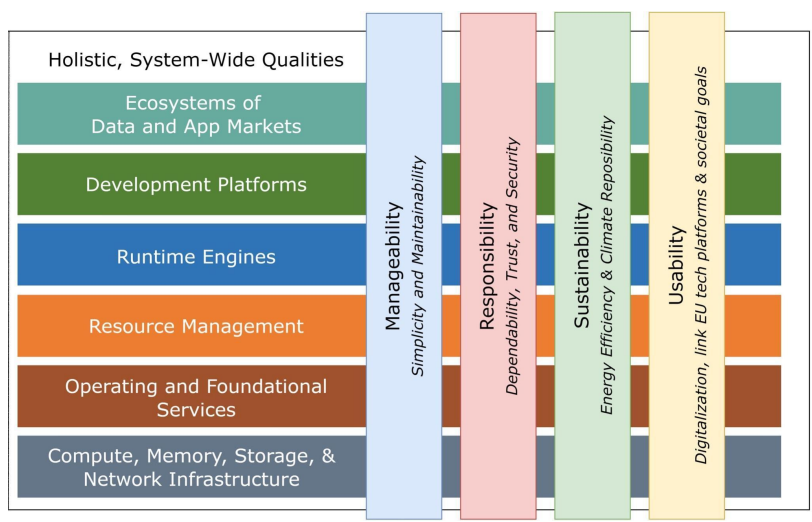

**Figure 1**. (Horizontals) Foundational research themes in computer science and networking research in the Netherlands, and (Verticals) Cross-disciplinary advances with outstanding impact on society.

# Executive summary in Dutch

Samenvatting van het manifest over Nederlands onderzoek naar toekomstige computersystemen en netwerken

**De stuwende kracht op weg naar een verantwoorde en duurzame samenleving? Computersystemen!** Onze moderne samenleving en concurrerende economie zijn afhankelijk van een sterke digitale basis en, daarom, van onderzoek en innovatie op het gebied van computersystemen. Computersystemen, variërend van kleine processoren binnen apparaten tot grote datacenters en de netwerken die ze verbinden, vormen een fascinerend technologiegebied met belangrijke impact op en kansen voor de samenleving. In dit manifest richten we ons met name op Nederland, waar datacenters al meer dan 3,3 miljoen banen en meer dan 60% van het BBP ondersteunen, nieuwe producten en diensten faciliteren, en waar elke €1 die wordt geïnvesteerd in computersystemen €15 aan toegevoegde waarde genereert. Door vroegtijdig te hebben geïnvesteerd in netwerkinfrastructuur is Nederland één van de grootste data-hubs ter wereld geworden en maken vele bedrijven en organisaties dankbaar gebruik van de cloud. Ons doel met dit manifest is om de grote maatschappelijke, technologische en wetenschappelijke kansen en uitdagingen in toekomstige computersystemen en netwerken onder de aandacht te brengen en te schetsen hoe Nederland een leidende positie op dit gebied kan behouden.

**Toekomstbestendige digitalisering vereist ICT-onderzoek en -ontwikkeling (R&D).** De Nederlandse overheid en maatschappelijke actoren onderschrijven het belang van ICT. Een verregaande digitalisering kan huidige activiteiten verbeteren, nieuwe disruptieve ontwikkelingen stimuleren en leiden tot hoge efficiëntie en, voor het eerst in de geschiedenis, universele toegang tot ICT. Maar zonder actie zou Nederland zijn prominente rol op het gebied van computersystemen kunnen verliezen, kunnen ICT-diensten hun flexibele, vertrouwde en veilige status binnen de digitale economie kwijtraken, riskeren we dat ICT-onderzoekers in Nederland vertrekken naar andere landen en zou het bestaande tekort aan ICT-geschoolde mensen alleen maar groeien. Onze belangrijkste bondgenoten, maar ook concurrenten, in de wereld, waaronder Duitsland, Frankrijk, de VS en China, intensiveren reeds hun langetermijninvesteringen in ICT R&D, met een sterke nadruk op computersystemen en netwerkinfrastructuur. En in Europa groeit het besef dat ICT ook soeverein en volgens Europese normen en waarden moet worden ingezet.

**Met dit manifest vestigen we de aandacht op computersystemen als vitaal onderdeel van ICT.** Computersystemen omvatten alle hardware en alle operationele softwarelagen die ICT toepassingen mogelijk maken. Elk van de topsectoren van de Nederlandse economie, elke route in de Nationale Wetenschapsagenda en elk van de Duurzame Ontwikkelingsdoelen van de VN poneren uitdagingen die niet kunnen worden aangepakt zonder baanbrekende vooruitgang in computersystemen en netwerken. Vooruitkijkend naar de periode 2030-2035, zullen er alleen belangrijke nieuwe diensten en toepassingen ontstaan als de ontwikkelingen in computersystemen dit mogelijk maken. Als gevolg van de COVID-19-pandemie zijn miljoenen mensen abrupt online gegaan, waardoor problemen met data-privacy en opschalen zijn ontstaan. Kunstmatige intelligentie (AI) vereist al enorme computersystemen, kan miljoenen kosten om te trainen, en heeft een onverantwoord grote energiehonger. Computerwetenschappen zoals bio-informatica, kunnen niet betaalbaar en efficiënt worden zonder een nieuwe generatie computersystemen. Evenzo is het opkomende kwantuminternet afhankelijk van (traditionele) computersystemen en netwerken. Grote commerciële sectoren, waaronder financiën en de maakindustrie, vereisen gespecialiseerde computers en netwerken. En veelbelovende technologie-hubs, havens en slimme steden riskeren afhankelijk te worden van niet-Europese technologie.

**Computersystemen moeten beheersbaar, verantwoord, duurzaam en bruikbaar zijn.** Om de potentie van onze gedigitaliseerde toekomst waar te maken en het risico van technologische afhankelijkheid te vermijden, moet Nederland *nu* vier grote maatschappelijke uitdagingen aanpakken met betrekking tot computersystemen:

1. **Beheersbaarheid**: Hoe moet de steeds groter wordende complexiteit, en de daarmee samenhangende menselijke fouten, in enorme computersystemen en netwerken getemd worden?
2. **Verantwoordelijkheid**: Hoe een verantwoorde computerinfrastructuur te realiseren – analoog aan het concept van verantwoordelijke AI – waarop we kunnen vertrouwen (vanuit het oogpunt van privacy, ethiek, beveiliging, prestatie, beschikbaarheid en duurzaamheid)?
3. **Duurzaamheid**: Hoe het energieverbruik van (toenemend) computergebruik te verminderen en optimaal benutten?
4. **Bruikbaarheid**: Hoe samen te werken, in het kader van de ambities van Europa aangaande digitalisering, met de voor de Nederlandse samenleving cruciale sectoren, zoals industrie, zorg en bestuur, om computersystemen voor iedereen bruikbaar en toegankelijk te maken? Hoe in de digitale infrastructurele basis voor AI en quantum te voorzien?

**Het aangaan van deze uitdagingen vraagt om een innovatieve, meerjarige visie en aanpak…**

**… door middel van een baanbrekende, holistische technologie-roadmap**
Nederland is goed gepositioneerd om een grote speler in computersystemen te zijn, vanwege zijn rijke historie op het gebied van communicatie en netwerken, en vanwege zijn huidige gemeenschap van onderzoekers op dat gebied. Omdat onze samenleving verregaande ambities heeft rond digitalisering, is de tijd rijp voor baanbrekende vooruitgang in alle lagen van computersystemen en netwerken. Denk aan de hardware, besturingssystemen, algoritmen om de werkbelasting te verdelen, ontwikkelingsplatforms, etc. Omdat de verscheidene aspecten van computersystemen onlosmakelijk met elkaar verbonden zijn, is het voor vooruitgang absoluut noodzakelijk dat men een holistische kijk op computersystemen en -netwerken neemt. Deze fundamentele onderwerpen vereisen dus ook een even fundamentele vooruitgang in de manier waarop we wetenschap bedrijven, namelijk richting holistisch systeem-denken en handelen.

**… door hernieuwd onderwijs, echt toegankelijk voor iedereen**
Als belangrijke bijdrage aan de Nederlandse samenleving hebben we een gecoördineerde actie nodig om systeemdenken opnieuw in te voeren als een duidelijk leerdoel, om zo onze samenleving te helpen de steeds meer onderling verbonden en complexe gedigitaliseerde wereld te begrijpen en vorm te geven. We hebben IT-professionals nodig, vanuit een brede vertegenwoordiging van de samenleving, met kernexpertise op het gebied van computersystemen.

**… en door een gezamenlijk antwoord op onze oproep tot actie**
In de race naar een digitale koppositie zien we een duidelijke kans voor Europa en Nederland om zich niet te concentreren op slechts één aspect van digitalisering, bijvoorbeeld AI-algoritmen, maar om holistisch de hele keten van computersystemen-gebaseerde technologie te beschouwen. Gemotiveerd door maatschappelijke behoeften en de sterke onderzoeksbasis op computersystemen en -netwerken in Nederland, roepen we op tot een grootschalige collectieve inspanning en investering op dit gebied. De voordelen van digitalisering zijn enorm en raken elk onderdeel van onze economie en samenleving. Doen we dit echter niet snel, dan is de kans groot dat zowel het computersystemen vakgebied alsmede de door onze samenleving gewenste ICT specialisten niet voldoende in Nederland beschikbaar zullen zijn. Economisch kan Nederland niet concurreren zonder computersystemen en netwerkonderzoek!

## Introduction

Computer systems and networks (CompSys) are the backbone of the digital infrastructure and, thus, are essential for all critical economic sectors and societal aspects of the Netherlands. Every bit of stored data and every computational task rely on a computer system to run on and, increasingly, on a network to reach it. A myriad of IoT and mobile devices produce vast streams of data, joining enterprise big data processes in using exabytes of data in the cloud or at the edge. In our modern digital economy, to "*out-compete is to out-compute*". The business environment of the Netherlands, an economy of nearly 1 trillion EUR per year [17], relies on computer systems and networks for daily operations, growth, and R&D. Data centers, which are enormous networked computer systems, support over 3.3 million jobs in the Netherlands and provide essential cloud services to the Dutch society during (rapid) digital transitions such as that caused by the COVID-19 pandemic; a staggering 60% of the GDP depends on digital operations running in Dutch and internationally operated data centers [7]. Computer systems operate decision-making and artificial intelligence-powered processes, collect and process big data, give access to performant and dependable cloud services, and make the digital infrastructure sustainable. The computer systems and networking field achieves this through diverse methods for creating, structuring, and organizing computing components and their communication networks, centering the components in well-behaved mechanisms and dynamic policies, using resource management, scheduling, and feedback loops, and enabling developers to interact with systems programmatically.

A responsible and sustainable future for the Netherlands relies on computer systems and networks. The list of sectors and domains that would benefit enormously from digitalization is broad, and includes: smart industry, intelligent healthcare, energy transition and energy markets, smart mobility and logistics, digital remote work, digital twins, digital entertainment and online gaming, digital government, digital education, digital culture, FinTech, smart agriculture, democratized and citizen science, the fourth paradigm of science (big science), and the emerging fields using AI and quantum computing.

Although the CompSys field has sustained remarkable developments since 1990, to achieve and improve digitalization it is imperative that our digital infrastructure keeps innovating. The arguments are pragmatic: Today, every €1 invested in computer systems generates €15 in added value for the Dutch economy, and innovation in computer systems, networks, and services hires an increasingly larger number of people, beyond the average economic sector [7]. However, the environmental cost is not cheap: Data centers powering cloud services consume about 3% of the national electricity budget, up from about 1% only 5 years ago, more and growing faster than in other countries (e.g., the US) and in the Netherlands significantly more than other infrastructure (e.g., for public transport [39]). With current technology, the amount of computer systems resources used for training AI systems is doubling every 3.5 months, threatening to become unsustainable and unecological [28]. Beyond mere numbers, the operation of government and societal services depends on increasing automation. Significant advances in CompSys technology are needed for each of the major routes in the National Research Agenda [39], each of the Top Sectors of the Dutch Economy [39], and each of the UN Sustainable Development Goals. Digital infrastructure is identified as an essential and thus far underrepresented part of the AI Research Agenda for the Netherlands [15]; AI systems cannot work without (new) computer systems and networks [28]. Similarly, the emerging quantum technology requires (traditional) computer systems to bootstrap operation [24], and should continue to act as a powerful complement rather than replacement for the next decades. These scenarios set the stage for **four grand societal challenges in the Netherlands** (the columns in the grid in Figure 1):

1. **Manageability**: Only two decades ago, connecting to the Internet and executing an application across multiple machines were grand challenges in CompSys; today, these are trivial tasks. The unprecedented capability of CompSys infrastructure led to a Cambrian explosion of new, heterogeneous technologies, addressing important societal demands but adding new challenges. However, to provide ICT services, all these technologies need to work together, often under the supervision of human experts. How to tame the ever-growing complexity, and related human error, in massive computer systems and networks?

2. **Responsibility**: As society is increasingly being digitalized, the importance of responsible CompSys is becoming paramount. Computing is shifting to a post-Moore's law world, where the principles change and performance advances cannot be taken for granted after nearly seven decades of rapid increase. In the last decade, scale and diversity have made CompSys infrastructures much less dependable, and security and privacy breaches have become so common that they can be now seen as the rule rather than the exception. Reliability and dependability are also advocated for software by VERSEN [53]. How to realize responsible CompSys infrastructure – analogous to the concept of responsible AI – whose operation we can trust and depend on? How to ensure privacy and security, along with performance, availability, and durability? How to enable ethical ICT for everyone?

3. **Sustainability**: "Fueled" by the European Green Deal, there are serious concerns regarding the energy footprint of (increasing levels of) digitalization. Other greenhouse emissions and consumption of limited resources are also major concerns. How to make the most from the energy footprint of computing? How to use electricity and resources most efficiently and how to offset their environmental cost? How to scale deployments elastically, on-demand? How to make use of efficient, power- and otherwise resource-constrained devices?

4. **Usability:** Critical sectors for the Dutch society, such as industry, healthcare, and government, need support to digitalize. People without digital skills or, worse, access to digital services could become effectively excluded from society. Questions of cross-disciplinary, translational science [30] arise, linked for example with the Dutch AI vision for an AI infrastructure [15]: How to work with these domains to make CompSys technology usable and accessible by all, and flexible enough to become valuable across all societal applications? How to test these properties, and gain valuable feedback for fundamental CompSys science? The Netherlands must also pursue **interoperation with and integration into a sovereign EU technology landscape:** Currently, our digital foundation is using mainly non-European technology, from Big Tech companies and countries that do not (need to) adhere to our data-privacy and societal goals. The European Union has started to react to this situation, with flagship R&D programs that aim to create core European technology, focusing on high-performance processors, high-speed Internet, energy-efficient and secure cloud, and sovereign data hubs. How to make Dutch R&D in computer systems and networks align, integrate, and interoperate with European technology platforms, and meet the EU societal and economic goals? Moreover, the grand goals related to AI and machine learning, and further in the future to quantum computing, depend on computer systems and networks as prime enabler and bootstrapping infrastructure.

We can overcome these grand societal challenges and transform them into great opportunities for the Dutch society. The Netherlands is well-positioned to digitalize society as a whole and maintain a competitive advantage. This is a consequence of its **rich, multi-decade tradition**, and of its current **community of CompSys researchers, whose productivity, quality, and vision are recognized world-wide** [6][12][13][18][19][27][41][46][51][52]. The Netherlands has a high-impact legacy and continues to play a prominent role in the development of computer systems and network technologies, from virtual laboratories for science to fintech and AI systems, from big data management to data hubs, from WiFi and Bluetooth communication to widespread Internet rollout. A thread ties history (for example, the Netherlands was the first European country to connect to the Internet) to present (for example, Amsterdam hosts one of the world's largest Internet Exchanges and the largest data hub in Europe, and various Big Tech companies are trying to establish here some of the largest data centers in Europe). These pioneering efforts have been a catalyst for the digital economy of the Netherlands, which today is among the best in the world, as for instance is illustrated in a report on the future of the Dutch digital economy [28].

However, that same report also indicates that *Dutch digital infrastructure is at risk of losing its top position* [28]. Worryingly, the Netherlands *may be entering a state of "digital disconnect":* Already, many aspects of the data centers and networks in the Netherlands are designed, manufactured, and even operated in non-EU countries, thus dwindling our digital sovereignty and draining critical Dutch expertise. Continuing to invest in these non-EU technologies may place the Netherlands in the unwanted situation where it has not aligned its digital strategy with Europe, yet it ends up needing access to the European infrastructure and digital market without being part of the research and decision-making processes underpinning this emerging infrastructure. The world has already started to up its commitment to new CompSys infrastructure, and in CompSys research: The US, Japan, and China, world-wide, and in Europe the main export-partners of the Netherlands, such as Germany, France, Belgium, and the UK, have made ever-larger investments to match increasingly more ambitious, ground-breaking goals in the last few years [29]. Small countries with outsized economies, such as Switzerland, have adopted similar policies. Large investments in AI systems have been matched by large (and even larger) investments in computer systems in all these countries.

Turning the tide requires a concerted effort that, inspired by our rich legacy, pushes the envelope in digital infrastructure through **foundational CompSys research**, **innovative education linked to CompSys research**, and **cross-disciplinary advances through translational research** [30] **in CompSys**. We formulate around these aspects an integrated vision in the field through **horizon 2035**. The **research and education priorities** identified in this vision require significant investment from the government and co-funding by industry, on-par with the benefits brought by successful CompSys technology and tolerating the risks of fundamental research; the Dutch Government needs to invest, because the local industry cannot do this alone, and the international industry has different priorities and may not invest in such disruptive technology. Moreover, other countries subsidize or otherwise fund their ICT infrastructure, possibly leading to an unfair advantage vs. the Dutch industry. Figure 1 links the foundational research themes in CompSys to cross-disciplinary advances with outstanding impact on society. We also formulate **general priorities**, such as a **coalition** of CompSys experts and users of (and in the Netherlands we have started to do so, through the IPN SIG in Future Computer and Network Systems[1]), and a **broad digital infrastructure** tailored to the needs of the Netherlands and accessible to the entire society. Because demands are varied, for example, computing needs in AI, big data, scientific computing, and business are very different from each other, the infrastructure should cover judiciously the full spectrum of computer systems, from High-Performance Computing (HPC) to cloud to edge to IoT, and from networks to beyond 5G communications.

To address the societal challenges, we identify **seven foundational research themes in CompSys** (the rows in the grid in Figure 1). They correspond to both historical and newly emerging foci among CompSys scientists, engineers, and other industry and societal stakeholders. They are driven by mainstay needs, such as high-performance processors, high-speed Internet, energy-efficient and secure cloud, and sovereign data hubs, and by novel paradigms, such as post-Moore and non-von Neumann hardware, the development of wireless, optical, edge, and beyond-5G services including low-latency and high data-rate, the moonshot of a transition toward a digitalized society launched by the Dutch Government [8], etc. They enable strong economic effects (see the box "Economic and societal impact of CompSys in the Netherlands"), to both established companies such as ASML, Adyen, and Booking.com, and to startup ecosystems that could enable the next break-through. But all these require overcoming the next research challenges. Uniquely, computer systems and networks are by definition composites, meaning the weakest part limits the capability of the entire system, so advances in one part at once reinforce and challenge the others. We identify **seven foundational research themes, which only when addressed together can ensure the four societal pillars are enabled**:

1. **Ecosystems of data and application markets**: designing and massivizing distributed ecosystems, that is, composites of interconnected (distributed) systems. Challenges:
   (1) creating processes that enable and facilitate pragmatic and innovative ecosystems, while focusing holistically on challenges including performance, scalability, availability, privacy, trust, security, (data) ownership, and ecosystem sustainability; (2) developing and calibrating approaches for (ethical) experimentation and design space exploration; and (3) developing the processes, tools, and standards to compose and deliver meaningful applications to empower digitalization and fulfill the grand societal challenges.

2. **Development platforms**: providing specialized tools for the development, testing, and deployment of (complex) applications that need to run over multiple machines and multiple tiers of devices. Challenges: (1) providing integrated development for the system as a whole; (2) dealing with system-level trade-offs; and (3) linking to software life-cycle, e.g., testing.

3. **Runtime engines**: translating digital applications into smaller units of computation easily understood and executed by modern computer systems. Challenges: (1) automate runtime engines without loss of performance, efficiency, security, etc., by scaling resources elastically and managing them; and (2) commodify them as services, by creating common, interoperable runtime engines that support different operational models and per-task/-user QoS needs.

4. **Resource management**: provisioning and allocating resources and services to ensure a specialized, distributed and parallel, environment. Challenges: designing algorithms and schedulers for fine-grained, QoS-driven (1) multi-tenant resource allocation, aggregation, and disaggregation; (2) workload and data orchestration, federation, and re-use.

5. **Operating systems and foundational services**: co-designing software and hardware for efficient operation within a single machine and over distributed setups. Challenges:
   (1) rethinking operating systems; and (2) designing flexible yet efficient multi-node (distributed) services.

6. **Hardware infrastructure**: providing compute, memory, storage, and network resources, both physical and virtual, for specific purposes. Challenges: (1) designing and developing the new generation of computing, networking, and sensing hardware, some general-purpose, some specialized for and co-designed with applications; (2) creating the foundations of virtual and in particular software-defined resources; (3) linking these resources to any deployment and environment.

7. **Holistic, system-wide qualities**: providing system-wide functionals and non-functionals, and specializing for particular goals. Challenges: (1) supporting performance, availability, reliability, security, energy-proportionality, stability, portability, and correctness; (2) enabling trustworthy operation; (3) ensuring simplicity, composability, and maintainability; (4) ensuring responsible operation; (5) ensuring energy-efficiency and, more generally, sustainability. Note the complementarity between this research challenge and some of the cross-disciplinary societal challenges, which provide context beyond CompSys for the (non-)functionals.

---

[1] IPN SIG FCNS: https://www.ictopen.nl/tracks/track-future-computer-and-network-systems/

The remainder of our manifesto proposes **a research and education agenda** along the seven foundational research themes, and further makes ten **recommendations** regarding computer systems and networking research, technology, and education in the Netherlands. We call for urgent and clear action to bolster the human and technical CompSys potential of the Netherlands. Our overarching goal is to help the Dutch society stakeholders realize the potential of CompSys research in the Netherlands, by creating today new opportunities that push the boundaries of what computers and networks will be able to achieve tomorrow, and building on this potential a modern, sustainable, and CompSys-empowered future. If we fail to do this soon, digitalization in the Netherlands will lack its backbone or, worse, sovereignty over it.

### Economic and societal impact of CompSys in the Netherlands

CompSys facilitates directly and indirectly many socio-economic activities in the Netherlands. Although the Central Bureau of Statistics (CBS) does not track distinctly and with fine grain the ICT, IT, and digital sectors, which are the main direct developers and beneficiaries of CompSys technology, several studies [36] assess the direct, indirect, and spillover effects of these sectors, in both economic and social terms.

Direct economic impact and job market growth:
1. In terms of direct impact on economic activity, "there can be no digital economy without digital infrastructure" [36]. The Dutch data center and cloud industry supports over 25% of the professionals in their daily work, a fraction which will exceed 35% by 2025 [36]. The total labor force at the end of Q4 of 2020 is 9.2 million, of which 8.9 million are employed.
2. The total Dutch Internet economy has an added value for the Dutch economy of over 110 billion EUR and supports over 330,000 jobs, of which 264,000 directly in ICT [36]. There are strong links with the European ICT sector and beyond, leading to large foreign investment in the Netherlands and various exports.
3. In 2019, the entire ICT infrastructure sector hired about 22,000 people, and produced €6.1 billion in total revenue and €3.1 billion as value added; by 2025, this should grow to 26,400 jobs (3% growth), €8.8 billion total revenue (6% yearly growth), and €4.5 billion value added (7% yearly growth) [36]. The data centers area alone created at least 12,800 jobs through 2020 and will provide 16,800 jobs by 2024 [36].
4. CBS estimates [36] the Dutch ICT sector yields an average gross value added of 5.2%, much higher than the average economic sector, which adds only 2.3%. Consequently, the field is growing with 5-6% yearly on average, and with nearly 10% for the ICT manufacturing area, which is much faster than the average sector.
5. Late-2020, CBS estimates [36] the total number of employees in the ICT sector exceeded 450,000, or about 5% of the total workforce (an increase of 9% since 2018), with over three-quarters having steady employment and 66,000 self-employing. These are well-paid jobs, as payment in ICT-intensive sectors is 18% more than in other sectors [36].

Indirect economic impact and job market growth, plus spillover effects:
1. CompSys sectors indirectly are expected to contribute to taxable, value-adding economic activity of €2.4 trillion (33% total added-value, rising) over the next 10 years, and impact economic activity of over €6 trillion over 10 years (60% of the GDP, rising), employing directly and indirectly over 3.3 million people [36].
2. In the Dutch digital economy, every €1 turnover in ICT infrastructure and services generates €15 in added value (projected €22 by 2025) [36]. This is much higher than the return on investment found for the US digital economy in 2018, where $1 invested in cloud services leads to $2.5 average net return on investment and a maximum of about $10 [36].
3. The added value of digitization and digitalized processes that characterize the next-generation digital economy in the Netherlands is estimated at €242 billion (33% total added-value) in 2019, and will increase to over €320 billion (40% total added-value) by 2025 [36]. This is faster growth than for other European economies or the US [36].
4. Pb7 estimates [36] the collective Dutch digital hub hires directly and indirectly about 39,000 people, with €4.8 billion added value. This is comparable with the physical Rotterdam Haven (largest impact, with 120,000 jobs and €14.5 billion added value) and Schiphol Airport

   (second-largest, with 93,000 jobs and €10 billion added value). However, the digital hub is growing faster (3-3.5x), has a much higher scalability ceiling through the 21st century, and already has an outsized indirect impact on the Dutch economy (2-3x larger).
5. CompSys has important spillover effects into GovTech, FinTech, energy transition, AI, and quantum computing, to name but a few immediate effects. For example, in the Netherlands, the GovTech industry is rapidly growing, with an addressable market on Justice and Safety of around €500 million, out of a total market for GovTech of €3.4 billion in 2021 [34]; in the EU, the GovTech industry is also growing, and has already reached €104 billion in 2021 [34].

Societal impact, through societal goods and services:
1. Inclusion and adult skills are greatly benefited by a generally available, non-discriminatory ICT infrastructure. Digital skills related to CompSys are essential skills for the workforce of the future. Access to ICT infrastructure for all should become a fundamental human right [18].
2. Many public services are moving online, including government online information services, online administration and taxation, digital signature and passporting, and even online politics. GovTech based on CompSys can help reduce administrative overload for public offices, and also facilitate access to such services in the general population, with less time, stress, and discrimination, easier detection of inconsistency or fraud, and easier application of rules and regulations.
3. Education, especially during the pandemic, is employing more digital services, which can facilitate sharing material, simplify academic accounting, and enable online access to experts and to sophisticated virtual labs. Services in health and social care, especially related to hospitals, could benefit in the same way.
4. The energy transition in the Netherlands depends on CompSys transition. Although this will also lead to significant economic opportunities and challenges, the climate impact achieved through this is a priceless societal service for future generations. Similar considerations apply to construction, agriculture and food production, water and urban management, and transport and mobility.
5. During the COVID-19 pandemic, ICT has acted as the main infrastructure for the general population, surpassing in economic importance any physical infrastructure. The return from this situation will likely not happen soon, if at all.

Summarizing, we argue in this manifesto for sustained investment in R&D in the CompSys area. CompSys is a fundamental enabler among ICT technologies, so commoditizing and outsourcing it greatly reduces the innovation and competitiveness potential of the Netherlands. Although investment in fundamental research is not enough to obtain societal benefits, the Netherlands has a strong ICT innovation ecosystem. Promising technology hubs, deltas, ports, and smart cities give innovation a chance for local (and global) success. The Netherlands has active university incubators, vibrant accelerators and crystallizers, and savvy (VC) investors.

## I. Research and Education Agenda

The strategic positioning of the Dutch CompSys community in the European and global landscape enable it to take a prominent role in addressing the critical challenges and opportunities that lie ahead in CompSys research and education, as outlined in this part.

Because CompSys thrives on combining many technologies into complex ecosystems, it takes time until a fundamental problem becomes noticed by the end-users of such systems -- in industry, governance, or society. A brief timeline exemplifies this. For many decades, computer systems leveraged inventions and trends from the 1960s, especially the microprocessor with complex architecture and Moore's economic law of exponential growth in the number of transistors per processor. These have enabled large gains in performance and cost, sometimes of several orders of magnitude larger than gains in other technologies and fields. But in the mid-2000s, these trends started to be overturned by the laws of physics and by the unsustainability of the approach into the future. Society became increasingly aware; for example, companies could not rely anymore on new devices offering large increases of performance over time (due to Moore's law). This led to a *paradigm shift*, based on sustained global investment in research and development, which has resulted in fundamental advances in computer systems and **a new generation of CompSys technology**: innovations in GPUs, TPUs, and other novel processors, distributed storage hardware and software, programmable networks, new data center stacks, and many others. Together with important interdisciplinary advances, such as in systems software engineering, systems resource management, and systems performance engineering, these ground-breaking scientific advances and methods will allow us to prototype and even deliver infrastructure with purpose and scale befitting the start of the 21st century. The Netherlands has strong competences in all these areas, with high-quality researchers located in Amsterdam, Delft, Eindhoven, Groningen, Leiden, and Twente, among others.

A trend of diversification beyond control also emerged in the 21st century. Whereas, since the 1970s, a narrow range of computer systems were available for a specific computing task and one could afford to learn a specific computer tool running on a specific computer and operating system, the present and especially the future are much more diverse. CompSys is currently taking a fundamental step, to the principles of *plurality and diversity* in (networked) computer systems technology, and more specifically to **computer and network ecosystems** where many different components developed for different purposes and by different teams evolve subject to relatively limited control of any single user; a related trend was observed for software engineering by VERSEN [53]. This will give the maximum opportunity for everyone in the society to benefit and take part in a fully digitalized society. However, this raises new challenges and opportunities that require high-quality research in computer and network ecosystems, which in turn will feed into our educational and professional development programmes. The Netherlands is well-positioned to become a leader also in these areas, especially through the clusters in Amsterdam, Delft, and Leiden, but needs a large investment in a national infrastructure for CompSys research to match this ambition.

Due to the many recent advances, education in CompSys technology currently has to cope with the complexity and diversity of topics without much help from traditional theory. Moreover, many problems are complex, non-intuitive, and require critical thinking to reach a solution. Thus, **research-oriented education in this area**, coupled with focus on experiments and empirical work, are staples of curricula that want to stay relevant. Because these aspects are costly and need refinement over time, and to **educate and train a truly diverse slice of the society in CompSys technology**, the Netherlands requires significant and sustained investments in CompSys education.

# 1. Priorities in Foundational CompSys Research

The time is ripe for groundbreaking advances in computer systems and networking. The CompSys researchers in the Netherlands are ideally positioned to offer leadership, and in particular in the last 5 years have provided (peer-reviewed) thought leadership for key topics: massivizing computer systems to achieve efficient, dependable, and sustainable computer ecosystems [18]; making a responsible Internet [13]; linking the cloud to edge into a *computational continuum* and pervasive data processing environment [41]; making the memory to storage continuum a reality [46]; co-designing hardware and software; interconnecting embedded systems, sensors, and IoT; enabling dynamic workload and resource management for the new societal needs and emerging workloads such as large-scale machine learning [52], graph processing [31][44], and online gaming [6]; providing full automation of ICT services and in particular concerns about infrastructure through serverless computing [31]; rethinking telemetry, monitoring, and real-time observation [50]; enabling programmability across all layers and components; seeking the principles of digital sovereignty and digital markets, etc. These fundamental topics require further, equally fundamental advances in how we do science in the field. We include here rethinking experimentation, simulation [19], reproducibility [51], and other essential methods; re-designing core processes for the performance engineering, software engineering, and design [19] of new computer systems; and re-linking the novel results into education together with new approaches to education in CompSys [20].

## 1.1 Ecosystems

> Our modern lifestyle depends on computer and network ecosystems, which combine all the other technology layers in Figure 1. The fundamental challenges include understanding, developing, and re-thinking the *integrated* technology; and also the processes and tools to engineer ecosystems aligned with EU goals and infrastructures, to enable computing and networking as a true utility.

As the backbone of the digital society, computer and network ecosystems provide the infrastructure for business, science, governance, and consumer applications. They create the necessary economies of scale that underpin broad participation and the innovation generating economic growth. Conceptually, ecosystems are composites of computer and network systems; and, recursively or as an overlap, of other ecosystems. Modern computer ecosystems exhibit a diversity of structure, components, and behavior that rivals biological ecosystems. Thus, **the main challenge is to integrate all these composites.** Fundamental research challenges here include: (**1**) establish the fundamental theory to predict ecosystem operation, dynamics, and evolution; (**2**) establish processes and develop tools to re-engineer key parts of existing ecosystems to gain observability and control, and ultimately sovereignty, over their operation in the spirit of the Dutch societal values and subject to EU and Dutch laws; (**3**) re-think ecosystems to meet by 2035 the key technical and societal aspects of manageability, responsibility, sustainability, and usability (the verticals in Figure 1); (**4**) by 2035, enable computing and networking ecosystems as the key utility they aim to be, providing techno-social and cyber-physical solutions across the **computational continuum**. As an orthogonal issue, (**5**) test the approaches empirically, systematically, and comprehensively. Table 1.1 summarizes the major topics, innovations, and technical terms.

The Netherlands is well positioned to play a leadership role in CompSys ecosystems in the world. Following pioneering work in distributed computing ecosystems in the past decade, the clusters in Amsterdam, Delft, and Leiden have established groundbreaking lines of research in the field, starting from the central premise that ecosystem dynamics and evolution, rather than the study of individual components in isolation, are essential to understanding and engineering ecosystems. Key contributions so far include establishing research programmes [13][18]; detailing some of the key problems tackled in these programmes [41][46][51][52]; establishing new design [19] and experimentation [27][51] processes for ecosystems, the last aligned with the ASCI vision for the NWO-funded DAS experimental platform; advancing resource management and (elastic) scheduling; increasing support for both general and specialized, and in general heterogeneous, resources and services; advancing capacity planning with a focus on long-term sustainability; establishing theoretical models (reference architectures) for emerging ecosystems, such as serverless computing [31]; considering fundamental aspects of reproducibility [51]; interoperating ecosystems; assessing and reducing operational risk; operating across computational continuum [46]; understanding availability and reliability of ecosystem services; uncovering emergent phenomena; understanding service use and adoption for various societal stakeholders [10]; developing and validating statistical and process-based models of ecosystem operations. These new lines of research should continue.

The open research lines aim to further address the four main societal challenges and include: enabling software-defined everything, including with humans in the loop; strengthening and expanding the role of non-functional requirements (NFRs, or the "how?" of the ecosystem) next to functionals (the "what?"); self-awareness supported by robust, explainable, and responsible AI/ML techniques; enabling super-distribution; new methodological approaches combining cross-disciplinary techniques from CompSys, software engineering, data management, and performance engineering; creating ecosystem visors, observers, and navigators for experts and the general public; creating reference models and predictive evolutionary theories to explain the formation and change of computer ecosystems [18]; etc. Applications of this research impact every aspect of the digitalization moonshot, including: efficient ICT-infrastructure management and operation, e.g., for a scalable, trusted, secure, and sustainable cloud; aligned with the Dutch AI Coalition (Nederlandse AI Coalitie) and the DutchAI vision [15], ecosystems for large-scale ML/DL with application in a variety of fields; aligned also with the SURF vision [45], flexible, scalable, dependable, and secure and ecosystems for science and big scientific computing, and for education; aligned with "the fourth paradigm for science", support for workflow- and data-driven computational sciences [14]; ICT infrastructure for democratizing science and enabling "citizen science"; pioneering computational models with a variety of applications, e.g., serverless computing [31] and graph processing [31]; national security (examples combine AI/ML and graph processing); future banking ecosystems; logistics; online gaming [6]; etc.

Researchers in the Netherlands already have access to basic research infrastructure to test their solutions, but the infrastructure for proper demonstration, proof-of-concept, and grand experimentation is currently lacking, thus posing clear threats to this field of study. Investment in this area, on-par with investments in the US (CloudLab, Chameleon sponsored by the government, environments shared by each Big Tech company), Germany, and other countries, is necessary. With this investment, the Netherlands could continue its participation in and links with EOSC, GEANT, ESFRI/SLICES, Gaia-X. On a small scale, the Dutch CompSys community has built elements of this infrastructure by complementing small seed-funding with its own equipment and volunteered administrative work, which indicates that researchers need and value research infrastructure. For example, the networking community joined under the 2STiC programme[2] and set up a nationwide test network consisting of programmable network devices, and the computer systems community formed the ASCI Doctoral School[3] and set up the nationwide distributed computing testbed DAS [2].

---

[2] https://2stic.nl
[3] https://asci.tudelft.nl/

**Table 1.1.** Summary of the **Ecosystems** research theme: (left) major topics, (middle) major innovations, and (right) key technical terms addressed by the research theme.

| Topics | Major innovations | Key technical terms in the field |
|---|---|---|
| Technology ecosystems | Massivizing computer systems; Responsible internet; Computational continuum | Complexity; short- and long-term dynamics; evolution; (unified) ecosystem theory; ecosystem design; ecosystem modeling and simulation; ecosystem experimentation; ecosystem engineering; ecosystem applications; ecosystem lifecycle; super-distribution; elastic scalability; NFRs; ecosystem resource management and scheduling; ecosystem heterogeneity; interoperability; energy-awareness and energy-efficiency; etc. |
| Business-oriented services | Digital economy; ICT-business value networks; Service ecosystems; Data Hubs | Service ecosystems and data hubs for verticals, e.g., smart industry (Industry 4.0), intelligent healthcare, digital energy sector and energy transition through ICT means, smart mobility and logistics, digital data and service marketplaces, digital remote work, digital twins, digital entertainment and online gaming, digital government, digital education, culture through digitalization, FinTech, smart agriculture, democratizing science, next-gen computational sciences, etc. |

## 1.2 Development Platforms

> Users interact with computer systems through a diverse portfolio of applications. The extraordinary increase in demand for applications - from digital entertainment and online gaming to personalized, intelligent healthcare and digital education, combined with the unprecedented sophistication of emergent functionality, mandates research into the design and implementation of effective and efficient development platforms, combining programming models and languages, libraries, compilers, and analysis tools, where productivity and flexibility play a central role.

User-diversity leads to a constant, fast-paced increase in the demand for novel, complex applications, often spanning multiple machines and multiple tiers of devices (including IoT). Without specialized tools, the development, testing, and deployment of such applications is simply too slow. At the same time, the technological push from the hardware and system architecture promises unprecedented performance and efficiency. Yet such emerging new technology must be made accessible to developers to facilitate the creation of novel applications that leverage these advances. Offering the right platforms to developers is the traditional domain of software engineering and programming languages, which are based, implicitly or explicitly, on specific programming models and abstractions, and substantial ecosystems of libraries and productivity tools that have emerged around them.

Development platforms must respond to the increased systems heterogeneity and high application complexity and diversity. To this end, **innovation on the development platforms is an urgent and inevitable necessity** in order to prevent a growing divide between hardware and software. Such innovation, however, must be guided by the human factor, as novel concepts and approaches must be made accessible to developers in the form of meaningful programming language constructs and effective tooling around them.

Addressing systems heterogeneity requires revising the design principles of development platforms in view of new emerging trade-offs between transparency, level of abstraction, and efficiency. The notion of transparency is a somewhat contentious topic in the area of development platforms, which becomes increasingly relevant again. Traditionally, coarse abstractions provide a perception of high usability and transparency, but gloss over the operational semantics of modern and sophisticated hardware without providing means for the programmers to interact with the hardware explicitly. As such, transparency may prevent efficiency and can, in some cases, even pose a threat to application correctness. For instance, the lack of explicit memory models in many traditional system-level programming languages like C has been the root cause for a plethora of difficult concurrency bugs, some of which only surface on specific hardware. Embedding explicit support for the relaxed memory models that modern hardware employs to facilitate better scalability of the chip designs towards larger numbers of compute cores is paramount to building robust and efficient applications. The same can be said for supporting more flexible concurrency models and asynchronous I/O operations to better align with how modern hardware functions.

To tackle application diversity and complexity, there is an increased focus on (distributed) programming at large**,** covering the edge-to-cloud continuum. Novel programming models and abstractions like serverless computing or Function as a Service promise high performance and efficiency for even fine-grained operations, expressing the coarse-grained parallelism required to scale large applications to the powerful and diverse infrastructure, but raise systems challenges even when managing small stateless functions [31], and more for general applications [31].

Thus, **the grand challenge** we face is **to design and implement integrated platforms that provide programmability and flexibility for applications and services development for the system as a whole.** To this end, the following fundamental challenges must be addressed:
**(1)** redesign abstractions for programming models to efficiently handle heterogeneity at multiple system layers, **(2)** design platforms with integrated support for large-scale, parallel and distributed programming, **(3)** design effective analysis/verification tools for aspects like correctness and performance, including AI-assisted programming and program synthesis, **(4)** support analytical and simulation-based models for portability and system-wide application prediction, and **(5)** provide tunable trade-offs between transparency, efficiency, and programmability, enabling users to choose the required levels of support, integration, and automation.

The Netherlands has a long tradition in designing development platforms, from compilers to programming languages. For example, research groups from VU Amsterdam, UvA, and CWI provide development platforms and tools for high-performance computing, machine learning, and embedded (real-time) systems, while research groups at TU Delft, TU/e, and U. Twente provide high-level programming models for specialized processors (like FPGAs and CGRAs), and groups at U. Groningen and U. Twente provide advanced analysis tools. However, the field is fragmented, and lacks the resources for a concerted research effort towards integration. We argue that tighter integration and wider adoption require fundamental research across all stakeholders - e.g., through a national effort towards providing the next-generation development platforms -, to prevent further fragmentation.

**Table 1.2.** Summary of the **Development platforms** research theme: (left) major topics, (middle) major innovations, and (right) key technical terms addressed by the research theme.

| Topics | Major innovations | Key technical terms in the field |
|---|---|---|
| Cloud computing | Serverless computing paradigm | Function as a Service, DevOps, pay-as-you-go computing |

| Programming models | Asynchronous computing and alternatives to thread-based concurrency | Futures, promises, coroutines, lightweight concurrency, actor model, lock-free data structures |
|---|---|---|
| | Dataflow-oriented models | Immutable data structures, ownership types |
| | Domain-specific languages | Portability, programmability, automated parallelization |
| Programming languages | Modern functional programming | Scala, functional constructs in mainstream languages like Java, Python, C++ |
| | Modern system programming languages | Go, Rust, explicit memory models, correctness-by-design |
| Tools and analysers | Integrated, system- and application-level tools | Compilers, profilers, concurrent debuggers, analysis tools, efficient verification |

### 1.3 Runtime Engines

> Runtime engines are the glue layer between modern applications and the computing, storage, and networking infrastructure. Runtime engines provide excellent opportunities for optimization and efficiency, but must respond to any significant changes in applications or infrastructure; this is a continuous challenge that, if lost, threatens to cancel out the benefits of any new technology - here, *lack of innovation is lack of benefit*.

Runtime engines use a variety of techniques to automate the use of compute, storage, and network infrastructure for specific applications. They simplify development by automating the use of various ICT resources and deployment by automating application-execution under various dynamic phenomena. They also provide critical feedback to both the application and the infrastructure operators. Runtime engines orchestrate the execution of complex applications, offering the developer sophisticated yet usable programming interfaces for advanced control. They coordinate with the Resource Management layer how to respond to the varying demands for ICT resources---e.g., by translating digital applications into smaller units of computation easily understood and executed by modern computer systems, or by elastically scaling the number of (virtualized or containerized) computing nodes as needed by the current phase of computation---; *this two-level coordination is essential for achieving efficiency in modern ICT infrastructure, and without two-level coordination there can be orders-of-magnitude more resource consumption and lower performance than with it*. Because innovation here depends on how emerging applications and infrastructure innovations will catch on, **this field is characterized by high-risk, high-reward developments.**

Currently, the runtime engine landscape is very diverse, with each application domain proposing engines to run specific application-functionality, e.g., Spark for big data, Tensorflow for AI/ML, Horovod for computation distribution, HDFS and HopsFS for storage, P4Runtime for P4-programmable networks, etc. Phenomena that emerged in the operation of modern computer systems, such as performance variability, impact of correlated failures, etc., have highly specific and non-linear impact on different application types. Thus, each runtime engine currently requires expert-knowledge to deploy, manage, and evolve. They involve one-time-off approaches, which are costly and error-prone, and not at all future-proof. Thus, **the current approach is not sustainable in the long run**.

**The grand challenge in runtime engines is to continue to innovate in its core operations, sustainably anticipating how emerging applications and ICT technologies will interact, and continuing to integrate with the Development Platforms and Resource Management layers**.

The fundamental research challenges here include: (**1**) innovate in execution engines, memory and storage engines, and network engines, trading off capabilities with expertise required to leverage them, (**2**) provide processes and tools to support the evolutionary process of migrating domain-specific capabilities to the Resource Management layer, simultaneously reducing the expert knowledge needed to deploy, manage, and evolve these capabilities, and, conversely, extending and adapting capabilities from the Resource Management layer for specific domains, and
(**3**) provide needed capabilities for observability, reporting, and fine-grained accounting, but also
(**4**) create processes and tools for common, interoperable, even cooperating runtime engines across applications and infrastructures [14], (**5**) offer support for emerging needs in emerging application domains (layer Development Platforms), especially as expressed in emerging programming models, abstractions, and interfaces, and (**6**) leverage the support offered by emerging capabilities and operational modes of emerging hardware and software services (layer Resource Management), especially as expressed in emerging resource management and scheduling processes. Approaches may overlap and integrate with those in neighboring layers, but even then will pose unique challenges due to the adaptation to specific application domains. For example, an ongoing challenge is (**7**) the transition of many domains to flow-based applications, which include workflows, dataflows, and control-flows (e.g., event-driven architectures commonly used in streaming), which across various domains have various degrees of, e.g., complex behavior, large scale, and dynamicity, and thus far do not allow for a general solution except for the simplest flows. Table 1.3 summarizes the major topics, innovations, and technical terms.

Addressing the needs of the digital society, application-developers create new applications at a rapid pace. The Netherlands is well-positioned to contribute and even lead in runtime engines for machine learning, big data and graph processing, blockchain, online gaming, and other business-critical and scientific computing workloads. The groups in Amsterdam, Delft, and Leiden are among the global leaders in FaaS and, more generally, runtime engines for serverless computing research, which aim to automate runtime engines and commodify them as services for universal use. The groups in Amsterdam and Leiden are global leaders in runtime engines for data-driven applications. The groups in Amsterdam, Delft, and Twente have key strategic roles in embedded, IoT, and industrial IoT research. Sustained investment is needed to expand this core expertise and make it a long-term driver of growth and digitalization capability for the Dutch society.

**Table 1.3.** Summary of the **Runtime Engines** research theme: (left) major topics, (middle) major innovations, and (right) key technical terms addressed by the research theme.

| Topics | Major innovations | Key technical terms in the field |
|---|---|---|
| Runtime engines | Powerful, self-aware runtime engines for specific application-domains | Support for emerging needs, across all engine-types, in machine learning, data analytics and graph processing, blockchains, etc. |
| | Runtime engine support for emerging hardware and services | Support for emerging hardware capabilities (accelerators, RDMA, etc.) and for emerging services (XaaS, FaaS, BaaS) |
| | Processes and tool | Support for sustainable innovation, toward common, standardized, interoperable runtime engines |
| | Execution engines | Automate and orchestrate: flows (workflows, dataflows, microservice-based); hosted and managed apps (some SaaS, FaaS, BaaS, some PaaS); 2nd-order and higher-order functions (Spark, MapReduce, Apache Flink/PACT); library-based distributed applications (MPI, Java RMI, CORBA, Akka); etc. |

|  | Memory and storage engines | Automate and orchestrate: file and data management; catalog and meta-data management; container and VM registries; active storage and ETL; key-value, key-map, file-base, and document-based storage; graph data management; object stores; schema and object-relational mapping; CDNs; programmable storage; consistency models; etc. |
|---|---|---|
|  | Network engines | Automate and orchestrate: data transport; object sharing; programmable networks; etc. |

### 1.4 Resource Management

> Resource management and scheduling (RM&S) is a key building block for any modern CompSys technology [13][18], automating most decisions about how to run the workload. RM&S has to ensure complex non-functional properties (performance, availability, security, energy-efficiency, all others expressed in the grand societal challenges in the Introduction). To do this, (future) RM&S has to be powerful, efficient, and self-aware, address both short- and long-term decisions uniformly, make decisions explainable to engineers, and keep the human in the loop.

Since the beginning of computer systems in the 1940s, RM&S has played an important role in managing the ICT infrastructure, in both the long- and the short-term. Initially a part of operations research and system analysis, the field has produced thousands of iterations over problems such as facility location (where to place computing hardware and how much, and where to interconnect facilities and with which networks) and workload scheduling (which part of the workload should run on which part of the infrastructure, when, and for how long). Hyper-specialized algorithms emerged, addressing complex optimization problems in offline and online conditions, able to consider multiple control variables and multiple objectives concurrently. However, the challenges faced so far have been a tame prelude for the **unprecedented scale, complexity, and dynamicity of today's infrastructure**. Workload, services, and resources need to be managed (across networks) to deliver the more traditional cloud Infrastructure-as-a-Service (IaaS), Platform-as-a-Service (PaaS), and Software-as-a-Service (SaaS), and more recently also Function-as-a-Service (FaaS) and Backend-as-a-Service (BaaS). New and diverse stakeholders pose new challenges, including common users who want to "plug-and-play", sophisticated end-users who demand on-demand infrastructure (e.g., network slicing [49]) subject to complex trade-offs, expert users who deploy business-critical Infrastructure-as-Code, COOs who want to operate as cost-efficiently as possible and thus demand elastic scalability, C-level officers who want to reduce operational risk and ensure against disasters; this leads to complex service level agreements (SLAs) that have become part of automated smart contracts that, when not fulfilled, trigger tangible and intangible penalties. Workloads need to combine scientific and engineering computing, with big data and AI/ML, and with network, web, and streaming capabilities (possibly real-time), all at various granularities, now commonly serviced by the same operator or even the same facility. The modern data center needs to operate a complex set of often thousands of support services, and combine heterogeneous hardware for both general-purpose and specialized computation, communication, and data. Data centers are disaggregated for security, fairness, and other purposes, but re-aggregated for performance, scalability, and yet other functional and non-functional objectives, and both as local and geo-distributed infrastructure; this poses new trade-offs. Fine-grained workload and resource management are now needed for serverless computing; containerized operations using Docker or Singularity, operating in Kubernetes-managed pods, complement traditional RM&S approaches based on Condor, Slurm, and Mesos. Already, cloud deployment needs to support complex models, including hybrid clouds, dynamic cloud bursting, and even full infrastructure federation. There are many complex emerging phenomena, such as performance interference and correlated failures, about which we know remarkably little; we should expect to discover more in the next decade. An increasingly stringent demand needs to be combined with all others: sustainability, including energy consumption (which must be minimized) but also energy production (with low $CO_2$- and GHG-emissions, linked to renewable production and other energy transition drivers) or "follow-the-sun" types of RM&S. Already, to take and enact their decisions, RM&S systems consume about 5% of the infrastructure resources [23].

**The main challenge is grand: To enable a new class of RM&S techniques, self-aware and able to co-exist without human interference, yet remaining under human control and becoming able to explain their decisions to ICT engineers**. Fundamental research challenges here include: (**1**) extend the RM&S system to its full complement of decision-making and -taking, both short- and long-term, and end-to-end, (**2**) enable powerful RM&S systems, based on artificial intelligence and simulation, (**3**) enable diverse feedback and machine learning loops, yet (**4**) keep the human in the loop, informed and with well-explained decisions, and (**5**) operate efficiently. As an orthogonal issue, (**6**) experiment with RM&S solutions systematically and comprehensively, preferably in properly sized real-world environments and/or with accurate simulations. Table 1.4 summarizes the major topics, innovations, and technical terms.

In the Netherlands we have long-term experience in RM&S, from operations research to CompSys-specific techniques. Researchers in every Dutch university have contributed to this field. Key contributions include all ranges and scales of the problem, from thought-formation on self-awareness and RM&S in general [13][18], to reference architectures and models for data center scheduling [1], to the development and tuning of detailed techniques for scheduling [16] and Quality-of-Service routing [25]. Our researchers have fundamental contributions to techniques for short-term RM&S such as load-balancing, replication, caching, partitioning, consolidation, migration, preemption, offloading, provisioning, allocation, elastic and auto-scaling, fault-tolerance, admission control, and workload throttling; and techniques for long-term RM&S such as capacity planning and disaster recovery [40]. An open research topic is how to fully and flexibly combine these techniques, at all temporal and spatial granularities, enabling a full complement of decision-making in a unified framework.

Self-awareness involves not only decision-making, but also sensing the reaction of the environment to the decisions and giving feedback on decisions. A recent proposal (authored by researchers in the Netherlands) details over 30 tasks related to the decision-making process [1]; another (led by researchers in the Netherlands), identified 10 classes of self-aware capabilities [21]; and a third provides a stepping-stone towards self-programming networks [42]. Extending the RM&S capabilities, powerful techniques are now available, due to much larger data and computation available for RM&S: for decision-making, *artificial intelligence and computer simulation (in silico experiments)*; for the feedback loop: demand estimation, workload prediction, performance prediction, and anomaly detection, based on *hand-crafted or machine learning models*. The various parts of the ICT infrastructure exhibit various degrees of these capabilities and become increasingly programmable and hence amenable to automatic RM&S; yet, the hundreds to thousands of decision points do not fully coordinate or may even not cooperate at all. Enabling self-awareness for RM&S, holistically over the various components, remains a grand challenge, for which researchers in the Netherlands are well-positioned to make significant contributions for the next decades.

Two topics have grown significantly in importance and require much more attention than in the past: the interaction between RM&S systems and their users, and the operational efficiency of RM&S. Both pose large and largely unaddressed challenges and require immediate attention from researchers. For example, enabling Infrastructure-as-Code and composable disaggregated infrastructure (CDI) are important goals that foster advanced usability. Decision-making should become more energy efficient.

**Table 1.4.** Summary of the **Research Management** research theme: (left) major topics, (middle) major innovations, and (right) key technical terms addressed by the research theme.

| Topics | Major innovations | Key technical terms in the field |
|---|---|---|
| RM&S | Powerful, self-aware RM&S | Artificial intelligence and simulation for decision-making; hand-crafted and machine learning models for the feedback loop; monitoring for RM&S |
| | Short- and long-term RM&S | Load-balancing, routing, slicing, replication, caching, partitioning, consolidation, migration, preemption, offloading, provisioning, allocation, elastic and auto-scaling, fault-tolerance, admission control, and workload throttling; capacity planning, disaster recovery |
| | Efficient RM&S | Energy-proportional decision-making |
| | Interaction users - RM&S | Infrastructure-as-Code; composable disaggregated infrastructure (CDI) |

### 1.5 Operating Systems and Foundational Services

Hardware innovation is advancing at a rapid pace, yet most software systems still assume an outdated, static hardware model. This disparity leads to decreased performance and efficiency, and also exposes the software to technological obsolescence as new hardware emerges that the software cannot use. Concurrently, important services, originating mostly in the runtime engine layer, are becoming common, even foundational services in the computational fabric. For society to leverage the power of modern hardware, a significant co-design of the operating system and foundational services (e.g., synchronization, messaging, consistent state management) is needed.

CompSys software has reaped the rewards of Moore's Law for several decades, building much needed digital services around CPUs that have grown constantly faster and more efficient. The fundamental abstractions that operating systems, storage systems, and distributed data processing systems are based on, however, are still largely inherited and have remained mostly unchanged from the 1980s. Besides the substantial technical debt they carry, there is also much room for improvement that can be gained by leveraging modern hardware, such as GPUs, FPGAs, TPUs, ASICs, and NVMe. Emerging IoT devices also require new, cross-layer hardware and software. To take advantage of these, we need to **rethink the core principles of operating systems**, and to **co-design system software and hardware for efficient operation for both single-machine and distributed setups**.

In modern computer architectures, the CPU is no longer the dominant element for modern demanding applications. Instead, we are now facing an increasingly diverse set of supplementary compute elements that are embedded into peripheral components such as accelerators, storage, network interface cards, and even memory. Traditional operating systems and foundational services still formally support these new devices through device drivers but lack the means to deliver their increasingly powerful capabilities to application developers, due to the lack of meaningful programming models and APIs. This leads to a sprawling landscape of vendor-proprietary solutions that all bypass the operating system and thereby undermine the portability of applications; worse, they lead to *vendor lock-in*. A prominent example is CUDA, the programming model for Nvidia's popular programmable GPUs, which is an irreplaceable component to making machine learning applications perform and scale.

The grand challenge for operating systems is to **overcome the gap in effective support for modern hardware**. Fundamental challenges here include: (**1**) re-shaping and re-envisioning how operating systems are layered and constructed, (**2**) finding efficient interfaces for applications and developers to interact with the foundational services, and (**3**) offering secure and efficient virtualization and resource sharing for modern hardware devices such as GPUs, FPGAs, and (programmable) ASICs, which, to date, have limited efficient and secure support for multi-user operation. Failure to address these challenges will put increasingly out of reach the traditional tasks of the operating system, i.e., abstraction, resource management, and safety (especially protection). As a result, it will be increasingly more difficult to develop applications that can operate efficiently on a variety of different computer systems, to ensure fairness and share resources between multiple applications running on the same specialized hardware, and to protect the already large attack surface offered by vendor-proprietary APIs that are impossible for experts to scrutinize and too different to harden efficiently. Table 1.5 summarizes the major topics, innovations, and technical terms.

The Netherlands has a rich legacy in operating systems, being one of the few places in the world where innovation in operating systems happened, e.g., through the development of Minix in Amsterdam. It is surprising how much knowledge and capability exists for this topic in the Netherlands, relative to the paltry funding made available for this topic. Unfortunately, some of the talent has started to drain to other countries, such as Germany and the USA. If we fail to act now, general operating systems, and more specifically knowledge in OS design, hardware, virtualization, and security, could become a thing of the past.

**Table 1.5.** Summary of the **Operating Systems and Foundational Services** research theme: (left) major topics, (middle) major innovations, and (right) key technical terms addressed by the research theme.

| Topics | Major innovations | Key technical terms in the field |
|---|---|---|
| Operating systems | Rethinking the OS | 21st-century hardware; massive multi-threading; asynchronous IO; isolation; containers; NVM file systems; fault models; persistent data structures (e.g., heaps); AI-driven data structures (e.g., learning indexes); network operating systems (SDN) |
| | Co-designing hardware and software | Heterogeneous computing; OS+hardware co-design; compilers for emerging architectures; runtimes for new languages and applications; software+hardware co-design; secure hardware+software |
| Foundational services | Multi-node services | Middleware; enterprise data bus (e.g., Kafka); distributed data services (e.g., file systems, key-value stores); distributed synchronization services (e.g., Zookeeper); cross-platform foundational edge, fog, and IoT services |

## 1.6 Hardware Infrastructure

CompSys infrastructure combines computing, storage, and networking hardware to support the large variety of modern (online) applications and services we use on a daily basis. The expectation for computing infrastructure is to continue to grow in capacity, capabilities, and efficiency

> indefinitely, constantly matching the increasing demands of the Dutch economy. To do so, especially post-Moore's law, modern infrastructure requires innovation in computing models and architectures, novel hardware designs, and systems architecture. There are also unique opportunities in cross-disciplinary research, with material sciences, neuroscience, or optical physics, contributing to achieving high-performance, highly efficient, and highly customizable CompSys infrastructure.

From the 1970s through the mid-2000s, the underlying hardware for CompSys infrastructure has converged to a rather limited set (e.g., general-purpose processing units and proprietary fixed-function devices) to facilitate large-scale deployment. Advances were centered around making these devices faster and more efficient, to achieve higher performance goals; improvements in CompSys hardware over this period were measured in orders of magnitude every few years, whereas in other technologies, for example in agriculture and transportation, improvements beyond 40-50% in a decade were considered remarkable. However, the end of Moore's law makes it increasingly challenging, and clearly not sustainable, to meet performance expectations using this traditional, exponential scaling. Thus, new CompSys infrastructure must rely on **new paradigms to address increased programmability and higher diversity.** Programmable hardware opens up the fixed-function boxes and reconsiders the allocation of responsibilities between hardware and software. It also reduces unnecessary features that otherwise are shipped by default. Higher diversity is traditionally achieved by specialized hardware and system (co-)design for domain-specific purposes, but these approaches remain to be validated for the new hardware paradigms that are emerging. In addition to these design challenges, research in hardware should take into consideration the current paradigm shift towards open hardware, which was boosted by the success of RISC-V. Furthermore, it is fundamental to guarantee the possibility to fabricate and test in realistic environments the new hardware solutions. **We predict and already have evidence that the era of one-size fits all is gone:** we need to design and develop efficient, hybrid computer systems by leveraging heterogeneous computing, networking, and storage devices with different performance, capabilities, and interfaces. We address each of these kinds of devices, in turn; Table 1.6 summarizes the major topics, innovations, and technical terms.

**Computing:** Processing power is often the flagship feature of computer systems. For many applications, like simulations in science and smart industry, video processing in digital entertainment, or massive data processing in training models for personalized healthcare, efficient execution – in terms of time, throughput, or energy – is crucial. Thus, these diverse and time-consuming applications have led to the development of new and diverse processing units. Today, computing no longer relies (only) on CPUs, but rather on a growing collection of heterogeneous hardware devices. For example, modern CPUs have evolved both in terms of instruction set architecture, with new incarnations of CISC and RISC solutions, and in terms of hardware organization, evolving from homogeneous shared-memory multi-core processors (like Intel's Xeon, AMD's Opteron, or IBM PowerPC families) to heterogeneous, hierarchical core organizations (like Cell B.E., Intel's Xeon Phi. AMD's Zen, ARM's BIG.little, or IBM Power9 families). The memory hierarchy has also seen a diversification of caches and scratch pad memories, both in terms of organization and technology.

Currently, the fastest, most impactful developments in the computing side of hardware also cover the realm of accelerators. Including GPUs, TPUs, DPUs, and, more recently, FPGAs, accelerators promise massive performance improvement (in the order of several TFLOPs) and energy efficiency. New hardware approaches, including optical neuromorphic computing and optical interconnects, could also reduce energy consumption and increase speed, but could require rethinking the hardware-software infrastructure layers. Similarly, in batteryless IoT, energy is harvested from the environment and therefore of intermittent supply, thus warranting a careful hardware-software design for computing and communication [54]. Nevertheless, the distributed nature of modern systems, and the tendency of sharing hardware resources, calls for novel design methods capable of ensuring the security of the computation. Programmable hardware, when supported by appropriated tools, offers a great opportunity to address current and future security requirements.

Finally, hardware design, and the implications of design decisions on the performance and efficiency of future computer systems, are more open and transparent than ever. Initiatives like Open Compute[4] enable users to truly oversee and rethink the core parts of processing units for applications/domains that require new specialized hardware for unprecedented performance or efficiency.

We foresee that future computer systems will become heterogeneous collections of diverse compute units [55]. This heterogeneity and the race toward exascale (and beyond [26]) create main challenges in **designing, building, and programming large-scale machines able to match different applications and user requirements**. Exposing and exploiting the specialized features of each of these accelerators requires well-designed communication and orchestration from both the system CPU and the system software. Thus, the fundamental scientific challenges to be addressed include: **(1)** co-design compute, storage, and networking for sustainable and/or predictable computer systems, even at exascale and beyond; **(2)** provide the right abstraction layers for specialized exposure of processing hardware at different software layers; **(3)** provide methods and tools for performance and energy consumption monitoring; **(4)** design composable performance models for processing units that enable applications to accurately budget for performance, energy, or utilization, thus reducing computing waste; and **(5)** provide tunable mechanisms, where trade-offs like performance vs security can be balanced in software.

The Netherlands has a strong presence in heterogeneous computing, with groups at TU Delft, UvA, VU Amsterdam, U. Twente, and TU/e designing specialized processors and accelerators for machine learning, real-time systems, and/or IoT installations. Large research projects like Efficient Deep Learning (EDL)[5], Energy Autonomous Systems for IoT (Zero)[6], and SMART Organ-on-Chip[7] feature hardware design and/or efficient usage, and showcase the proficiency and competency of Dutch academia to be at the forefront of such developments. We argue that a reduced role in the international development of hardware is insufficient for sustainable CompSys research in the Netherlands. Therefore, we need to increase the contribution from the Netherlands in programs like the European Processor Initiative, where currently only SURF participates from the main Dutch knowledge landscape. Companies producing hardware, software, and ICT services could gain a more prominent role, in collaboration with universities and with a stimulus from the Dutch government.

**Networking:** Since its advent in the 1960s, the Internet has grown from a handful to billions of nodes. Internet services have expanded from sending email to conducting online business, education, and entertainment. The Internet is built with a suite of standardized network protocols which allow the network to be built in a decentralized manner and to scale to large installations. To match growing network traffic demands, current practice relies on incremental updates: more devices (with higher capacities) are added. Predominantly, we rely on devices with fixed functions running the protocols.

However, the proliferation of modern online services; the pursuit of higher data-rates, reliability, scalability, and lower latencies; and the increasing risk of security breaches and unintentional human error call for a diverse and possibly dynamically changing set of network functionalities, in addition to simple traffic forwarding. With fixed-function devices, network functionalities are manually implemented and wired into the network. This process is tedious and error-prone, when considering the number and the variety of devices, and their complex dependencies. The fast penetration of 5G networks and associated brand-new network services indicates such an approach will soon become unsustainable.

---

[4] https://www.opencompute.org
[5] https://efficientdeeplearning.nl/
[6] https://www.zero-program.nl/
[7] https://bit.ly/SMART-OoC

Software defined networking (SDN) aims to address this challenge by introducing more programmability in the network architecture. SDN allows separating control from the data plane, and using software and possibly artificial intelligence techniques to control the network. This enables, for example, enhanced orchestration. Also, the data-plane has seen the emergence of novel programmable hardware architectures like Protocol Independent Switch Architecture (PISA) and programmable network devices (e.g., Tofino switches, SmartNICs, NPUs) that can be customized with programming languages like P4 and NPL.

Broadly speaking, in addition to advancing the physical layers (optical and wireless), and leveraging technologies like chiplets and silicon photonics to improve the networking stack, **the main challenge for networking hardware is to achieve true programmability and seamless system-level integration without trading in on speed,** thus turning the fixed-function devices running protocols into a programmable network system where network functionalities can be specified with software on demand. To achieve this goal, we must address the following scientific challenges: **(1)** define modern high-level programming models/abstractions for network devices, contributing to the integrated development platforms (see Section I.1.3), **(2)** design and prototype new hardware architectures and devices, such as programmable photonic network devices and data-plane programmable beyond 5G type of communications, **(3)** realize end-to-end programmable and convergent photonic-wireless networks, **(4)** propose methods and mechanisms for network configuration synthesis, network automation, network slicing [49], and intelligent management [4][48] of network resources, and **(5)** investigate new security risks and work towards security-by-design for programmable network devices and (open) technology.

**Storage:** Storage systems are a critical part of computer systems and services. Historically, disk capacity and packaging densities have improved rapidly over time, while bandwidths and access latencies lagged. Since the mid-2000s, we have witnessed the rise of storage based on Non-Volatile Memory (NVM). NVM storage technologies such as Flash or Optane can nowadays support commodity, mass-storage computing with up to 10s of GB/s of bandwidth, single digit microseconds of access latencies, and millions of operations/s, rivaling DRAM speeds. These innovations represent a fundamental shift in the computing hardware and storage architecture, towards the development of a unified memory and storage technology. Not only has the performance improved significantly, but the devices themselves have gained capabilities to support multiple higher-level software features such as programmability, scheduling, virtualization, security, and efficient data management.

**Table 1.6** Summary of the **Hardware Infrastructure** research theme: (left) major topics, (middle) major innovations, and (right) key technical terms addressed by the research theme.

| Topics | Major innovations | Key technical terms in the field |
|---|---|---|
| Compute | Domain-specific accelerators, computing models | Free and open Instruction Set Architectures (RISC-V), GPUs, TPUs, Neural Engines, FPGAs, ASICs, neuromorphic computing; battery-free computing; performance monitoring and modeling; computation-in-memory |
| Networking | Programmable networks, new network architectures, advanced | Programmable photonic network devices, photonic-wireless convergence, THz communications, Visual Light |

| | physical-layer technology | Communications (VLC), Software Defined Networking (SDN), Network Function Virtualization (NFV), programmable data-plane |
|---|---|---|
| Storage | Programmable storage, virtualization support, novel I/O operations, persistent memories, unified storage-memory computing models | NVMe, Zoned Namespaces, Optane Data center memory, Project Silica (from Microsoft), SNIA working group specification on programmable storage, application-specific storage architectures, memory-centric/in-memory comp. |
| All | Use of innovative materials | Advanced materials, optical physics, neuroscience: glass and resistive memories; photonic networks; etc. |

Looking forward, the **grand challenge in storage systems is to combine heterogeneous storage layers, leveraging their programmability and capabilities to deliver a new class of cost, data, and performance efficiency for all kinds of applications**. This requires rethinking the entire storage hierarchy, from hardware to applications and services, to design hybrid, flexible, and active storage services, which further enable hardware/software co-design. The fundamental research challenges to reach this goal are: **(1)** re-designing storage APIs and abstractions to efficiently leverage storage device heterogeneity and programmability, **(2)** co-designing processing units for seamless integration with different layers of storage, **(3)** integrating storage with the network, aiming towards a unified storage-memory-compute and computation-in-memory(-stack) model, **(4)** co-designing mechanism for multi-tenant secure storage systems through resource isolation, **(5)** investigating emerging storage technologies (e.g., DNA, glass storage, and carbon nanotubes).

## 1.7 System-wide Qualities, e.g., Performance, Availability, Energy-efficiency, Privacy, Trust, and Security

> Modern computer systems need to work well, which poses necessary compromises (trade-offs) between performance, availability, energy consumption, privacy, trust, security, and other non-functional requirements. Among all scientific and technical communities, the computer systems and networking community has the unique capability to address the fundamental challenges in expressing, optimizing, and explaining these 21st-century trade-offs. This is because these aspects need to be addressed systematically and holistically.

Users demand increasingly more from modern computer systems and networks: it is no longer sufficient for an application to simply work; how well it works, where it is executing, or how well it protects each user's data are requirements that cannot be ignored anymore. These and other *non-functional requirements* impact the design, development, and deployment of computer systems. Experience shows that such non-functional requirements are often interdependent, in what we call *trade-offs* – such as application speed vs. energy consumption. Trade-offs dictate the way we build and use computer systems today but will continue to affect society well into the future, because the ICT infrastructure has real physical presence and climate impact.

With increased digitalization, systems and applications have become significantly more complex, and, therefore, more complex trade-offs emerge; users are increasingly expecting that applications are simultaneously responsive, accurate, and available, governments are starting to become aware of the need to limit the climate footprint incurred by ICT infrastructure, industry demands that ICT operations are secure and resistant to attacks from outside the national borders, and society is beginning to demand that ICT infrastructure is trustworthy, privacy-preserving, and usable for all. Not enough understanding exists about these novel needs and related trade-offs. Consequently, no current approach can robustly optimize for all these aspects.

Novel components and design methodologies are needed, in particular, for security. Weaknesses and attacks that exploit information leakage (such as physical attacks or microarchitectural side-channel attacks) have demonstrated that system security can be achieved only if it is considered since the beginning of the whole design phase, and it is tackled holistically. Current approaches that often consider security only at the last stages of the design must thus be replaced by methods to construct systems that are secure and privacy preserving by design. And while security is a *conditio sine qua non* for trust, it is not the only aspect. Trust also relates to transparency and control – can users know and exert control over how their data are handled – reliability, and more.

**The grand challenge here is to make system-wide qualities easy to observe, reason about, and manage optimally.** Fundamental research challenges here include: (**1**) holistic support for expressing, optimizing, and explaining different trade-offs, including through real-time, digital twins of critical infrastructure; (**2**) enabling Operational Data Analytics (ODA) [33] across the entire ICT infrastructure; (**3**) security-, trust-, and privacy-aware computing and communications (by design); (**4**) processes and tools to support a transition to energy- and climate-aware ICT infrastructure; (**5**) FAIR data and open-source software [32] for responsible science and societal access to knowledge and decision-making in this field; and (**6**) processes and tools to explain these trade-offs to all societal stakeholders. Table 1.7 summarizes the major topics, innovations, and technical terms.

**Table 1.7.** Summary of the research theme on **System-wide Qualities**: (left) major topics, (middle) major innovations, and (right) key technical terms addressed by the research theme.

| Topics | Major innovations | Key technical terms in the field |
|---|---|---|
| General NFR, including: | Holistic support for expressing, optimizing, and explaining 21st century trade-offs; Responsible explanation to all societal stakeholders | Support for complex trade-offs, such as performance-availability-cost, performance-trust, and security-performance-energy; selection by the user/app; actionable efficiency analysis; co-design applications and infrastructure for one trade-off; digital twins for operational decisions; holistic analysis and modeling of operations and risk |
| (1) performance and dependability | Operational Data Analytics (ODA) | Collect and FAIR-share fine-grained operational data; FAIR open-source software for ODA analytics; enable proactive and reactive feedback loops; enable proactive and reactive analysis, control, and optimization loops; cross-disciplinary research for processes and tools to visualize ODA results and to use them in the control rooms of ICT infrastructure |
| (2) security, trust, and privacy | Trusted computing and communications (by design) | Tools and methods for secure and open hardware; processes and tools for intrusion/anomaly detection; processes and tools for technology benchmarking and auditing in terms of security, trust, and privacy; ensuring security, trust, and privacy by design; trusted and secure hardware platforms; responsible internet |
| (3) energy efficiency and sustainability | Novel, energy- and climate-aware computational and execution models | Transition from mere energy-efficiency to climate-awareness; energy- and climate-aware infrastructure; processes and tools for energy analysis and modeling; approximate computing with energy constraints; energy- and climate-aware DevOps; cross-disciplinary research to achieve smart energy generation, transport, and use in ICT infrastructure |

The Netherlands is well-positioned in this area. Research groups across the country have focused on complex optimization across various trade-offs and diverse classes of ICT infrastructure. Research groups in Amsterdam and Delft have pioneered empirical research methods to collect and analyze ODA data for various classes of ICT infrastructure, and are global leaders in this field. Amsterdam, Delft, Nijmegen, and Twente host world-class researchers in cybersecurity, trust, and privacy, pioneering aspects related to digital independence and Internet security for the past decade and currently working together to realize **the responsible internet** [13], that is, a vision for an internet that harbors users' privacy and enables enhanced transparency and control, by leveraging open and programmable network technology. Research groups in Amsterdam, Delft, and Twente are actively developing processes and tools related to energy and climate; and are aligned with the emerging LEAP leadership-class alliance in the Netherlands. Aligned with the leadership role played by the Netherlands in establishing the FAIR principles [32], a cluster of high-quality researchers, located in Amsterdam, Delft, Leiden, and Twente, has collected and shared FAIR data and open-source software related to system-wide qualities, for over a decade.

*These strengths of the Dutch research community in this area are real, but to keep up the field requires immediate investment on par with the benefits society derives from ICT infrastructure, similarly to the large growth fund (Groeifonds) investments made recently in other digitalization processes, e.g., in AI algorithms and systems*. We are still at the moment when we can address the grand challenge, before the construction of too many hyperscale data centers and related ICT infrastructure, and before the security and energy-consumption flaws become too large. Failure to do so will mean outsourcing optimization and losing sovereignty over the results, further eroding our (digital) independence. More subtly, failure to do so will impose tough choices on future generations: live with indispensable ICT infrastructure that is not fit to address 21st century challenges and that exposes society to large risks, or a costly rebuild after enormous expensive infrastructure has already been deployed.

## 2. Cross-Disciplinary Translational CompSys Research

Multi- and interdisciplinary research, in short, cross-disciplinary research, runs the risk that the CompSys component of this research is used only for its applied research value, thus both limiting its potential contribution to the development of the other sciences involved in the process and diminishing its ability to learn back from these other sciences. Instead, we advocate for *cross-disciplinary translational research in CompSys*, with three core pillars: (**1**) experimental work focusing on grand experiments and on the ability to "break" seemingly working CompSys artifacts under very specific conditions; (**2**) experimental work focusing on use cases, proofs of concept, and pilot trials that help understand and cross-pollinate with the needs of another discipline; and (**3**) co-design processes and best-practices in collaboration with professionals from other disciplines, aimed to promote and enable digitalization through CompSys technologies.

The main difference between translational research in CompSys and the current applied research is that *translational research uses bi-directional exchange of knowledge and best-practices*. We are inspired in this approach by recent advances in translational research in computer science [30], which are in turn inspired by the biomedical field; translational medicine aims to "combine disciplines, resources, expertise, and techniques within these pillars to promote enhancements in prevention, diagnosis, and therapies" [30]. This applies across all Technological Readiness Levels (TRLs) used by EU-level projects.

### 2.1. Simplicity, Composability, and Maintainability

> Digitalization is not possible without taming the complexity of computer systems and networks, by prioritizing simplicity, composability, and maintainability. These aspects need to be considered with a cross-disciplinary perspective, where CompSys plays a prominent role alongside AI, and contributions from experts in law and regulation, policy making, economic ecosystems, ethics, and processes involving the general population (e.g., through GovTech).

A study by ENISA regarding Telecom Security Incidents 2020 [8] reveals that the vast majority of telecommunications incidents are not caused by malicious actions, but rather by system failures and human error. These are clearly CompSys aspects and point to the need to tame the ever-growing complexity in computer systems and networks[8]. Particularly the use of AI techniques may aid operators in controlling these ecosystems; the nascent AIops field aims to combine CompSys and AI techniques to enable this.

As Turing Award winner[9] Edsger W. Dijkstra put it, "Simplicity is prerequisite for reliability". Unfortunately, partly due to their success, computer systems and networks are becoming more complex by the day. For example, when we look at networks (but the same trend applies for CompSys in general), it is not only the sheer size of networks, but also the diversity of protocols and algorithms that exist to govern how our data traffic is routed. And the specifications of some of those protocols can span hundreds of pages. Moreover, each router vendor is using its own vendor-specific language, which must be mastered to configure the protocols. And given that there are many protocols, each with many configuration settings, we end up with near-infinite configuration possibilities. This complexity, as indicated by Dijkstra's quote, is the reason why misconfigurations are one of the main causes of network outages. This complexity also stands in the way of the Dutch and European pursuit of digital sovereignty. Much of today's telecommunications operations are outsourced to non-European countries. To regain control is not only a matter of cost, but also requires sufficient in-house expertise and personnel.

---

[8] Other fields have had to overcome the complexity gap while delivering real-time decision-making and control, e.g., aircraft piloting for passenger, freight, military, and other applications. Solutions in these fields are never one-size-fits-all, but nuance and holistic approaches for specific purposes.
[9] The Turing Award is the computer-science equivalent of the Nobel prize.

Partly due to the aforementioned complexity, often an approach is adopted where certain building blocks of the CompSys ecosystem (like protocols) are designed in isolation. As a result, when deployed "in the wild" alongside other protocols and algorithms, undesirable effects, such as unfair resource utilization or high-performance variability, do manifest [47][51]. To avoid such undesirable effects, a holistic approach to monitoring and configuration is needed, but this is simply too complex to do by hand. In other words, the operator would benefit from a higher degree of automation, verification, and possibly auto-completion of configurations in computer systems and networks [42][43], which in turn requires new research. Similar research is also actively pursued in the software engineering community. For example, in June 2021, GitHub announced *copilot*, which uses AI to help programmers write better code.

These technical aspects need to be complemented by cross-disciplinary translational research with experts in law and regulation, policy making, economic ecosystems, ethics, processes involving the general population, etc. GovTech is an emerging (and already lucrative) area [34] aiming to provide socio-technical solutions for digitalized processes and digital interaction in the public sector, typically developed and delivered through public-private partnership. The **main cross-disciplinary challenge is to protect public values** such as *social inclusiveness* especially regarding access to CompSys-mediated technology for the elderly, new immigrants, and other social categories that may have trouble with the complexity of digitalization; *support for expected values including privacy and tolerance*; *reduction of bureaucracy, corruption, and general stress of administration* through CompSys-mediated solutions that "just work"; ability to *combat fraud and other digital threats, and more generally to defend the principles of security,* through CompSys-mediated processes; etc. Together, these are necessary components of any working *democracy*.

## 2.2. Responsibility, Trust, and Security

> Even if accomplished technically, digitalization must enable long-term activities based on responsible, trustworthy, and secure processes. CompSys experts should therefore engage with these topics, in collaboration with experts from related fields. Responsibility, trust, and security are cross-borders, so the Netherlands should collaborate on the European level and beyond.

The recent ransomware hack of a major US oil pipeline [11] is unfortunately but one example of how disruptive computer and network outages can be to society. It is evident that we should protect society against such disruptions, which calls not only for **embedding security by design, a grand challenge on its own, but also enabling aspects like transparency** (can we know how our data is routed and processed?)**, digital sovereignty** (can we have control over it?)**, and responsibility** (who is accountable?).

Our computer and network systems have become too big and important to fail, yet they frequently do (for various, including malicious, reasons). It is also the case that, in key areas such as governance and education, the Netherlands is one of the large consumers of cloud services hosted by cloud operators whose headquarters are not in the Netherlands or even in Europe, much more so than other countries in Europe [10]. Both aspects relate to trustworthiness, i.e., do we trust the system to consistently provide the services we need, and do we trust that our precious data are not misused by the companies offering those services. Especially when those companies are based outside of Europe, our digital sovereignty is at stake, which leads us to the following challenge: how to extend the original set of design goals for (our own) computer and network systems with concepts like robustness against side channel attacks and security-by-design, transparency, reliability, trustworthiness, etc., and to enable verifiable insights into and control over the operation of those systems? Furthermore, certain computation could be carried out without the actual need of sharing, at least directly, the data. However, the practical use of these constructions is currently impossible, because they can require too much computational effort. Many of the design goals, e.g., for the Internet area, are open problems since decades, so to add the aforementioned design principles, research is needed into: (**1**) novel metrics and measurements to assess/verify the state of the systems and the privacy of the data used in the computation, (**2**) mechanisms to balance transparency and security objectives, (**3**) new hardware paradigms, including (open) programmable hardware, or radically new approaches to use existing ones, to effectuate near-instantaneous control, (**4**) languages for expressing trust and sovereignty requirements, etc. and end-to-end over all compute, storage, and network resources used, (**5**) novel approaches that allow to ensure trust and to track the flow of information in a practical settings, (**6**) approaches for carrying out computation while maintaining the privacy of the data (such as homomorphic encryption) and their trade-offs with other (non-functional) goals.

On a European scale, the Gaia-X project aims to develop a data infrastructure based on such values of openness, transparency, and trust. The Gaia-X project focuses mostly on providing cloud services and less on making the network itself also more transparent. This is where the recently launched Dutch project Controllable, Accountable, Transparent: the Responsible Internet (CATRIN) comes in. CATRIN proposes two network planes: (**1**) The Network Inspection Plane (NIP), which makes transparent to users which network operators (e.g., ISPs or DNS and cloud providers) handle their data flows, and whether those actors align with the users' principles. (**2**) The Network Control Plane (NCP), which, based on insights gained from the NIP, empowers users to have a say in how the Internet infrastructure should handle their data.

Similarly to Section 2.1, the link with GovTech is strong, posing cross-disciplinary challenges in
(**1**) technological transparency for CompSys and CompSys-mediated solutions, which increases *trust* in governance and GovTech in general; (**2**) trust-enabling technology including in the CompSys layers, with additional trade-offs between privacy and law-enforcement capabilities, and further ethical and societal discussion; (**3**) efficiency of secure solutions, which should be measured at least against CompSys performance, scalability, and energy consumption; etc. This way, we can ensure that the people of the Netherlands have the right to use ICT, to learn how to use it, and to understand their own use as it happens [18].

### 2.3. Energy Efficiency, Climate Responsibility, and Sustainability

> To tackle globe climate and environment-related challenges, the EU has set the goal of achieving no net emission of greenhouse gasses in 2050 and decoupling the economic growth from resource use in the European Green Deal [9]. As a major energy consumer, the ICT domain shares this responsibility and requires a deep reform towards energy awareness to meet this goal. This includes both coming up with more energy-efficient designs for computer and network systems and algorithms, and involving more renewable energy sources in provisioning ICT infrastructures.

Our living environment is surrounded by computing technologies, from small, embedded sensors on our home appliances to mega data centers powering up sophisticated online services. Yet, all these technologies require electricity to function. As suggested by a 2020 meta-analysis of available data [35], ICT infrastructure accounts for more than 2% of global electricity consumption; in the Netherlands, CBS reports [35] an even higher fraction, of 2.6% in 2019, but growth since 2017 is faster than elsewhere [35]. With the rapid rollout of 5G networks and the Internet-of-Things (IoT), the number of computing and networking devices keeps growing drastically [3]. As a result, drastically reducing the energy consumption of cellular communications is one of the main pillars of the vision for 6G. Meanwhile, the ever-increasing data volume together with the fast penetration of artificial intelligence in various industries and business domains have increased the computational intensity on these devices to an unprecedented level [3]. The digitalization of our modern society will keep pushing these trends. At this pace, it is predicted that by 2025, the ICT sector will consume over 20% of the entire world's electricity [35] ; in the Netherlands, the total electricity consumption will become larger by a factor of 2 up to 7 by 2030 [35], over the 2019 consumption of 2.7 TWh [35].

Overall, we are facing the following **grand challenge: How to make our computer and network infrastructures energy-efficient and environment-friendly, while sustaining the continuous growth?** Fundamental research challenges associated with this grand change are: (**1**) new hardware technologies that can provide similar performance and capabilities, but with much reduced power consumption, (**2**) new computing models and architectures (e.g., neuromorphic computing) that can perform computations more efficiently, (**3**) energy-efficient system designs and software for runtime systems and resource management, (**4**) hardware-software co-design with built-in energy efficiency, and (**5**) integration of renewable energy in computing and networking infrastructures, and harvesting energy from the environment to power batteryless IoT devices, to make them truly energy- and climate-aware, and thus sustainable. An alternative to using less energy for the same work (challenges 1-4) and to using greener energy for the same work (challenge 5) is to do less work altogether: (**6**) finding technical and societal approaches that lead to reducing the amount of work - for example, caching and using approximate computing, and incentivizing professionals to skip running the least useful workloads, respectively.

The Netherlands has a long history of using renewable energy and is well positioned to address the above challenges. In particular, the Dutch government has put tremendous effort into energy-efficient ICT with initiatives like the GreenICT foundation, SDG Nederland, and Green IT Amsterdam. A special focus on data centers has been put, since data centers have been the backbone of our digital society and the Netherlands has a strong and mature economy around data centers. On the research frontier, the NWO-funded Perspectief projects like ZERO and EDL have been tackling the energy-related challenges in systems for IoT and deep learning, respectively, which are among the most important application domains today.

### 2.4. Digitalization in the Netherlands and aligned with the EU

> The transition to digital processes is happening in all key sectors of the Dutch economy and is a key goal of the Government of the Netherlands. CompSys is a cornerstone of this transition, and through translational CompSys research in collaboration with experts from each sector (and application domain) the transition will continue successfully. CompSys can help harmonize national needs and European Union goals and policies, both economic and societal.

**The grand challenge for digitalization in the Netherlands is that technology and processes must be tailored to the national needs**. Challenges include: (**1**) translational CompSys research to support the digital transition; (**2**) translational CompSys research to provide digital data and service marketplaces (linked to Sections 2.1-2.3); (**3**) translational CompSys research to provide specialized support for *industry verticals*, e.g., for smart industry, intelligent healthcare, a digital energy sector, energy transition through ICT means, smart mobility and logistics, digital remote work, digital twins, digital entertainment and online gaming, digital government, digital education, digital culture, GovTech, FinTech, smart agriculture, democratized and citizen science, the Fourth Paradigm of Science (big science), Quantum computing, and other domains.

**A further grand challenge** derives from the geo-political situation of the Netherlands: **CompSys in the Netherlands must align and integrate with the EU technology platforms, and help meet the economic and societal goals of the EU**. Challenges and opportunities derive from the alignment with and integration into the EU strategy, specifically, (**1**) digitalization in general: EU single digital market; (**2**) CompSys technology in general: the EU related strategy for green and digital technology, EuroHPC, the future 5G Toolbox for wireless communication in Europe, and the future European Open Science Cloud (EOSC) and Gaia-X; (**3**) focus on trust, privacy, and security: New EU Cybersecurity Strategy, EU General Data Protection Regulation (GDPR); (**4**) specific domains: New Industrial Strategy, Pharmaceutical Strategy for Europe, Europe's Beating Cancer Plan; (**5**) climate change: European Green Deal; (**6**) focus on science: European Research Area, including Horizon Europe; (**7**) focus on the assessment framework: EU Digital Economy and Society Index (DESI).

The community has built several research infrastructures to test their theoretical constructs and practical prototypes, especially in Amsterdam and Delft, and in collaboration with organizations such as Deltares and governance, or through fieldlabs such as the Do IoT fieldlab[10] in Delft, etc., but more investment is needed to acquire the large-scale infrastructure needed for proof-of-concept tests and grand experimentation.

---

[10] https://doiotfieldlab.tudelftcampus.nl

## 3. Priorities in Research-Inspired Education in CompSys

We highlight the need for IT professionals with core CompSys expertise, coming from all backgrounds and including currently under-represented groups in the field. We need to further consider societal and peopleware [18] problems in largely CompSys-oriented courses, and train students from all walks of life to recognize CompSys-related problems. Challenges in education also include didactic elements: because computer systems courses are both broad and deep, we need to help our students engage and stay motivated, and give them exploration tools to put theory to practice [e.g., 27].

### 3.1. Teach Computer Systems Thinking in All Relevant Curricula

> Currently, advances in most science, technology, and information studies rely heavily on computing and networking. As such, computer systems are ubiquitous in each of these fields, but often invisible to users. This lack of awareness and understanding of computer systems leads to their inefficient use in terms of performance, security, and interoperability; more subtly, users should develop an intuition about computer systems. To eradicate such problems, all curricula in these studies should include computer systems knowledge to some degree, ranging from basic to advanced according to need. It is only with such knowledge that true systems thinking can be taught and applied in modern science and technology, and in society at large.

A digitalised society relies heavily on computing and network infrastructure and services. To further advance and even accelerate the development of the Dutch digitalised society, the next generation of users, developers, designers, and integrators must be aware of the role of computer systems in all these aspects of our lives. To raise this awareness, several aspects of the education curricula at different levels must be adjusted to include age-appropriate and skill-appropriate notions from computer systems. Failure to adapt the curricula to include such notions will have profound negative effects in terms of using, designing, and building digital applications and services. In contrast, successful integration will likely attract more and diverse specialists in a domain that often struggles to recruit home-grown talent. Importantly, but more subtly, both specialists and the general public should develop an intuition about what computer systems can do and how they can do it, especially regarding efficient, sustainable, and secure computation and communications; this is much like how scientists and engineers have to develop an intuition about statistics or analytical modeling.

Current education curricula shy away from topics related to computer systems, which are perceived as (overly) technical. Thus, computer systems challenges and opportunities become invisible for our next generation of students, specialists, and scientists. This process must be reversed, or we risk outsourcing a lot of our digital services to companies elsewhere, thus losing governance and control over these services. Already today the demand from companies for (local) CompSys personnel surpasses the availability.

We advocate for **a unified effort to support computer systems modules in relevant curricula at all education levels**. Integrating computer systems in all relevant curricula, nation-wide, requires defining several levels of proficiency, from basic to expert. Concretely, we propose the development of the following modules: (**1**) basic computer systems literacy for everyone in a digitalised society,
(**2**) fundamentals of computer systems functionality for students in scientific domains,
(**3**) fundamental principles and operational skills for students at universities of applied sciences, and
(**4**) basic design, fundamental principles, and operational skills for students of technical universities and computer-science or related fields. (We address experimental CompSys topics in Section 3.2, 'Teach Performance and [...] Experimental Research in CompSys'.)

Additionally, we believe that enabling a digitalised society in which no one is left behind requires strong support for **national programs for life-long learning through training**. Computer systems literacy and knowledge at different levels (basic, proficient, advanced, and expert) can and should be facilitated through dedicated learning modules and activities. The computer systems community can design and maintain such educational learning modules, with the logistic and financial support from universities and governing bodies, such as the Ministry of Education, Culture, and Science (MinOCW).

From a **didactical perspective**, CompSys also presents a diverse palette of opportunities to hone specific skills in computer systems thinking, from theory to experimentation. These include, for example: (**1**) project-based education, where students could develop their research skills by going through a complete research cycle, from the formulation of a research question, to generating and analyzing the results, to reporting about it; (**2**) the use of (social) gamification in education, where aspects from (online) gaming are introduced in complex courses, e.g., on distributed computing systems [22], to boost student engagement; and (**3**) Massive Open Online Courses (MOOCS) and other forms of online learning (much used during the COVID-19 pandemic), where the field of CompSys itself also directly enriched the way we teach.

## 3.2. Teach Performance and Other Non-Functional Aspects, Benchmarking, and Empirical and Experimental Research in CompSys

> Consumers of digital applications and services primarily are often perceived as caring only about application functionality. However, non-functional properties – like execution speed, efficiency, energy consumption, reliability, security – are becoming equally important, due to their significant impact on the overall sustainability of and trust in our digital economy. All creators of digital content must be aware of these non-functional properties and should assume an active role in overall efforts to build more sustainable systems. This starts with their education: all education targeting digital content creation must contribute to raising awareness and improving the state-of-the-art in terms of non-functional properties for applications and services. Hands-on practice, on realistic infrastructure, can play an important role in forming intuitions about real issues.

Designing applications often starts from functional requirements: a definition of what the application must or should do. However, in current design and development processes, additional (and often, much stricter) requirements dictate *how* an application must perform its tasks. These non-functional requirements often refer to minimizing execution speed or energy consumption, or maximizing the efficiency in utilizing resources or security, and contribute to the overall sustainability of a digitalised society. Ignoring such requirements often becomes problematic only at deployment time, when, for example, a neural network training or a digital twin simulation become too slow to handle; worse, some of these requirements, such as the amount of electricity consumed by the application, are not even visible to the user. Thus, these problems are often (wrongly) associated with and blamed on computer systems, and the solution is, invariably, to buy or build bigger systems – for example, buy a larger machine or a faster network. In turn, this often leads to poor resource utilization[11] and a significant setback in efforts to increase the sustainability of ICT.

We argue instead that **the problem is the lack of awareness about non-functional properties**: what their impact is, how they can be measured and monitored, and how to account for them in digital content creation, from scientific simulations to visualization tools, and from games to financial calculations. Thus, all higher-education curricula where digital content creation is part of the exit qualifications and/or skills share the responsibility of raising awareness about these non-functional properties.

---

[11] Historically, server utilization has been below 15%. Many modern clouds have not increased this value beyond 50%. In contrast, well-managed infrastructure uses over 90% of its resources.

To this end, we propose to **design dedicated teaching modules** for (**1**) introducing the basic concepts behind non-functional properties of applications and computer systems, and (**2**) showcasing empirical research methods and tools for monitoring these non-functional properties in real applications and systems. Teaching the fundamental concepts will facilitate, for future domain experts, the transition from considering mainly functional aspects (i.e., "what operations should be performed?") to a grounded attitude, where applications and systems are to be understood also non-functionally (i.e., "how well are the operations performed, and at what cost?"). Teaching empirical methods and tools will facilitate the transition from thinking in terms of predictive theory to the reality that, at realistic levels of complexity, most systems act very differently from expectation. We envision such modules being designed and developed together with domain specialists (scientists, AI experts, simulation experts), and diversified to target different academic levels: BSc, MSc, and PhD students of applied science, technical, and general universities.

A type of experimentation very common in professional CompSys is **benchmarking**, which allows comparing multiple approaches with standardized, reproducible procedures. Teaching and training using benchmarks could develop important intuitions and help deepen the material, and emphasize the importance of reproducibility. However, this type of education also requires **adequate experimental infrastructure and instrumentation**, which for CompSys purposes can be expensive in comparison with typical laboratory equipment needed for computer science. Fieldlabs, in addition to serving as a place for research and innovation, may also serve as one vehicle to unlock realistic infrastructure for students and professionals.

### 3.3. Teach Co-creation with Other Domains

> With computing at the core of research and development in many domains, all key sectors of the Dutch economy, and all routes highlighted by the NWA ("nationale wetenschapsagenda"), it is easy to assume a producer-consumer relationship, where CompSys provides the computational power and tools to be used by other domains. This approach fundamentally limits the pace of progress and development in both the consumer and the producer domains. Instead, we advocate cross-disciplinary research, where both domain and computer scientists collaborate to co-design effective solutions, and where feedback is returned to all sciences involved. To enable this paradigm shift, tools and methods for this new kind of research must be taught at MSc and PhD level.

Virtually all scientific discoveries in the past decade have relied on computing and, by extension, on computer systems. For example, advances in radio astronomy, genome sequencing, natural language processing, and automated translation have all relied on smart algorithms running on increasingly powerful computer systems. Nobel and Spinoza prize winners can easily point to the CompSys infrastructure they relied on and, notably, include CompSys experts in their award-winning thanks. Similarly, the industry relies on computing and networking in a wide range of domains, from financial calculation to product simulations to sensing and control in smart industry settings. Emerging domains with bright economic and societal future, such as AI, GovTech, quantum computing, rely on CompSys expertise; for example, quantum computing currently relies on traditional computer systems to simulate its revolutionary techniques and algorithms and on classical networks to set up the entanglement for the quantum internet [24].

Traditionally, computer systems are seen as the infrastructure that supports the idea, and not as a part of the idea itself. In effect, this means a disjoint development of these applications from computer systems, where domain experts and CompSys experts interact in a producer-consumer relationship. Such a disjoint approach has two major drawbacks: (**1**) it slows down development in both areas, as alignment of requirements and solutions is often expensive in both time and resources, and (**2**) it cancels the premise of cross-disciplinary research that could lead to truly innovative, effective, and efficient research instruments dedicated to the task at hand. The latter is more subtle, but a common situation is currently that scientists from an emerging domain lose an early start, because they cannot envision what CompSys infrastructure could do with envisionable developments, and CompSys scientists and technicians take a long period of time before they can help, because they do not understand where the science aims to go and what are its challenges.

To avoid these drawbacks, we posit **it is essential to teach cross-disciplinary translational research in all fields, including CompSys** – CompSys science and other domains are intertwined and should not be separated. In this way, effective co-design approaches and methods can emerge, ultimately fostering an effective collaboration between all interested stakeholders. Such an ambitious goal is not without challenges: (**1**) devising a cross-disciplinary translational research education program that is widely accepted in different domains, including CompSys, relies on widespread literacy in CompSys concepts and research methods, and (**2**) establishing cooperation and support from diverse academic and research bodies, including investment in pilots for designing and/or improving methods and tools for co-designing effective and efficient research instruments, and obtaining buy-in and commitment from university staff and students, to embrace such a curriculum.

### 3.4. Teach about ICT-Related Societal Problems for a Healthy Society

> The rapid increase in connected devices per person (estimated at 4 to 6 per person by 2023 [3]) is illustrative of the massive impact and opportunities for CompSys worldwide. However, these opportunities come at a steep price, including enormous energy consumption, security crises, information leaks or censorship, and ethically questionable data-analysis. We must ensure our education curricula openly present and discuss these challenges, and monitor the approaches and solutions. In turn, this will open a clear path towards a healthier digital society.

Although with great benefits, such as to vaccine design and development, or to understanding and mitigating climate change, CompSys does not come cheap: energy consumption for all our devices, from personal and IoT devices to supercomputers and data centers, is steadily increasing. The impact of this growth on the energy and other resources of the world, and, consequently, on climate change and further on the long-term sustainability of humankind, can no longer be ignored.

Moreover, the countless opportunities for CompSys research and applications have led to an explosion of systems and tools that are, very often, open to the entire world. Together with the massive amounts of data generated by all of us online, and through networks the accessibility of compute resources in cloud installations, these applications and systems can be easily abused (maliciously or through lack of awareness). For example, open-source frameworks and just a couple of remote compute nodes enable almost everyone to train a massive neural network to recognize faces in the crowd or to generate deep-fakes. Distributed Denial-of-Service attacks can be executed at the click of a button, by simply purchasing them on the dark web. Social networks, with all their benefits and challenges, are enabled by CompSys ecosystems. And in particular in the area of online gaming and online social media, toxic behavior has been growing along with the popularity of CompSys devices. And while clear toxic behavior should be mitigated, there are also plenty of examples where censorship is misused to oppress. These are but few examples to indicate that ethical, legal, and regulatory aspects need to be addressed as well, next to the technical challenges raised by ICT infrastructure and services.

All these issues can pose significant threats to our well-being as individuals and in society. Therefore, **the CompSys community must openly discuss these challenges and rally behind the common goal of contributing solutions to these challenges.** For it to have an impact, this discussion must include the next generations of CompSys experts and users. We argue therefore that all the aspects related to the socio-economic impact of CompSys, from reproducible to comparable quantitative information, from sustainability to security, and from ethics to economics and law, must be included in all relevant curricula, at academic level. To this end, however, a concerted effort is needed to build and maintain a credible, fact- and numbers-based analysis of all these threats, their impact, and potential solutions. Paraphrasing David J.C. MacKay[12], we need to discuss "CompSys, without the hot air".

### 3.5. Teach Creative and Design Thinking for Young Specialists

> Massive leaps in the development of CompSys often arise, much like in many other domains of science and technology, from blue-sky, curiosity-driven research, and from the goal to push the boundaries of current systems beyond their known limits. Therefore, the search for the next generation of CompSys must rely on young specialists with the knowledge and tools for creative and design thinking. We advocate that advanced CompSys education must invest adequate time and resources, and make room for creative and design thinking in the curriculum.

Many great discoveries in CompSys – time-sharing, the PLATO system, the Internet, the multi-core processor, GPU computing, or the computational ecosystem behind AlphaGo or the CERN Large Hadron Collider – have started as creative solutions from blue-sky, curiosity-driven research, and resulted in paradigm shifts. To ensure such discoveries will keep emerging, we must **encourage our young CompSys specialists to think big, but also equip them with the tools and methods for creative and design thinking**.

To reach this goal, we envision **specialized education in creative and systems design thinking in CompSys**, with five core principles: (**1**) teach and challenge the fundamental design principles for computer and network systems and their evolution in time, (**2**) provide actual snapshots and examples of both successes and failures, and emphasize that history is not a long string of successes but rather an exploration between solutions that often take time to separate, (**3**) present and discuss the fundamental challenges, technological and societal, that CompSys had, has, and will still have to solve, and avoid the simple turn of phrase "this is old technology", (**4**) provide evidence of creative, cross-disciplinary solutions that have led to paradigm shifts in CompSys and beyond, and (**5**) present methods, tools, and processes [e.g., [19]] to facilitate CompSys design, and help the transition from CompSys design to implementation to realization to operation. These aspects contrast deeply with the descriptive approaches, and even to the analytical ones, that are commonly taught today.

This vision, however, is not immediately applicable. For example, while the history of CompSys is increasingly better documented, an open challenge is building (and collectively agreeing upon) the genealogical trees (more likely, graphs) of the various systems that led to significant paradigm shifts or to significant failures. A different, yet equally difficult challenge is selecting the short list of inspiring cross-disciplinary solutions that showcase the process and impact of creative thinking in CompSys; likely, scientific and networked cloud ecosystems will be a source. Also of great significance is the challenge of defining and agreeing upon the language and methodology for CompSys design, such that its separation, especially from implementation and operation can be clearly explained.

As CompSys is a rapidly evolving domain, this content must be maintained and updated. We recommend this specialist education to start from an online manual, adjusted for the specific curriculum, to which the CompSys community can constantly contribute, as for example initiated by the computer networking community [5]. Annual revisions and a lean consortium of research and education representatives to manage its development could address this issue. Imagine the young specialists that will create the future, equipped with the right cognitive tools of the day!

---

[12] The seminal book "Sustainable energy, without the hot air", by the late MacKay, is relevant also for CompSys and has been made available for free online: https://www.withouthotair.com/.

## II. General Priorities and Recommendations

> The Netherlands needs a roadmap focusing on investment in CompSys R&D, changing the current approach of mere (or just late) adoption. Societal stakeholders should value innovation *in* computer systems instead of merely *with* computer systems, in developing cross-disciplinary translational research in CompSys, and in high-quality education and training coupled with diversity goals. A key focus of these recommendations is the holistic view on CompSys research; all of it is reinforced by advances in any of it.

The Netherlands has strong research in computer systems, but more investment and organization is needed to maximize its strengths and potential, preventing a brain-drain that is already starting.

We make 10 recommendations, focusing on the priorities of CompSys in the Netherlands. Each of these recommendations is high-reward, and due to mutual reinforcement combined they could lead to major improvements for the Netherlands. They can lead to pragmatic innovation in CompSys, through and beyond horizon-2035.

### R1. Invest in CompSys research personnel in the Netherlands: a structural investment of €195 million in people, co-financed over the next 10 years by Government and industry stakeholders, to deliver its holistic view.

CompSys technology is the cornerstone of the Dutch digital economy and has an outsized impact on the Dutch economy and society[13]. For the next 10 years, ICT will have an estimated direct contribution of over €1 trillion in added value, indirect contribution of over €6 trillion, and will support through spillover the creation of new sectors related to GovTech, AI, quantum computing, and others. Moreover, the technology push and societal pull of the 21st-century strongly indicate that current CompSys technology is about to be replaced with a radically new generation. Early investors will not only create new goods and services before other advanced economies, but also reap the benefits and establish themselves as mediators in the global digital economy.

We recommend that investment in CompSys research in the Netherlands becomes proportional with the operational cost of ICT infrastructure, and linked with the impact it provides on the economy. The Ministries of Economic Affairs and Climate (EZK), Internal Affairs (BZK), and Justice and Security (J&V), have important interests related to this field [37] and thus should play a direct role in developing CompSys research in the Netherlands. Their Nederlandse Digitaliseringsstrategie 2018-2021 is a good step in the right direction, but significant investments in CompSys research are still to appear.

**Our pragmatic recommendation is to invest and fund over the next decade in the order of 500 PhD students, and academic experts and support staff to help coordinate and develop these new experts.** This would make the program similar to, for example, the large investment in Sweden in AI systems, and the estimated final investment in the Dutch equivalent *groeifondsproject* NedAI. The main technical outcome of this investment is a proof-of-concept CompSys platform that takes a holistic view of the challenges and opportunities at every level in the technology stack that we envision as the economic workhorse of the Netherlands in 2035-2040; in other words, we will leverage all the advantages offered by the current generation of academic experts in the Netherlands, as described in Section 1, 'Priorities in Foundational CompSys Research'. The economy able to absorb and

---

[13] See our box, 'Economic and societal impact of CompSys in the Netherlands'.

leverage this expertise already exists; however, a structural investment in it, for example through the *Groeifondsprogramma*, could deliver the very positive scenario in which **the Netherlands will maintain the attractiveness of its global data-port through 2035 and beyond** (akin to the physical Rotterdamse Haven and Schiphol Luchthaven), and will **bootstrap an unmatched, high-quality digital ecosystem around it, with Dutch and EU policies and goals**.

This significant investment in personnel should be matched by **proportional investments in experimental infrastructure and in a network of excellence, see next recommendations**.

### R2. Establish and maintain a coalition for CompSys, through a direct investment of €15 million over the next 10 years.

The complexity and diversity of CompSys ecosystems makes it an area of research that is addressed by researchers from various disciplines (see Section I.2, `[Cross-Disciplinary Translational CompSys Research](...)`'). Guaranteeing well-performing CompSys technology thus warrants that those researchers unite and collaborate towards shaping the future of computer systems and networks. Within the Netherlands, we have provided a platform for doing this through the newly established Special Interest Group (SIG), on "Future Computer and Network Systems," of the ICT-research Platform Netherlands. Presently, this SIG has representatives from most Dutch universities, and its focus is primarily on research and education related to CompSys.

The long-term goal of this SIG, and a recommendation from this manifesto, is to **establish a larger CompSys coalition, of which the IPN SIG would be a part, but which also has members from industry and government**. Given the importance of responsible digitalization, such a collaboration in triple-helix format is needed to assure that (**1**) the right research problems are addressed, (**2**) results are more easily adopted in practice, and (**3**) relevant governance aspects are developed concurrently. This could be extended to directly include representatives of the Dutch public, e.g., through citizen-scientists, and through NGOs interested in the societal and climate impact of CompSys.

To achieve these goals, the coalition needs to bring together and coordinate all parties covering the entire triple helix, improving cooperation and information sharing across stakeholders, etc. Pragmatically, this requires operational and administrative staff, and a direct investment proportional with the ambition. We recommend an initial direct investment to create this structure, plus periodic investments to (**1**) support specific actions aimed at stimulating large research activities and public-private partnerships, (**2**) offer overviews and operational insight into the operation of CompSys infrastructure, and (**3**) inform the general public through specialized and general events. Several leading candidates already exist to establish and maintain the coalition, but essential funding is currently lacking.

This recommendation is linked with the significant investment in personnel (R1) and in research infrastructure (R3).

### R3. Invest in CompSys *infrastructure* for research, demonstrations, proof-of-concepts, and pilots: a structural investment of €145 million over the next 10 years.

As our society has become increasingly digitalized, computer systems and networks *are* being used on a daily basis. This provides a clear opportunity to put theory, e.g., the development of algorithms and mechanisms, to practice, and to use the insights gained from those practical deployments to further tune the theoretical contributions. Indeed, **CompSys is an experimental science** [2][18].

However, infrastructure for computer systems and networks is expensive to procure, and for some areas there is also the need to support the fabrication of novel experimental hardware or the development of new specialized software. The cost of supercomputers and backbone routers runs in the millions of Euros, for example, the experimental supercomputer Piz Daint (Switzerland) costs around €35 million only to keep up-to-date, and the operational costs for 2-3 years match the cost of the initial hardware and software. Such costs are often out of the range of researchers to (develop,) purchase and experiment on. There exist some initiatives in the Netherlands, like the DAS cluster (several interconnected clusters with advanced processors, network, and storage devices) and the 2STiC testbed (consisting of a small nation-wide test network of programmable devices), but these initiatives often focus on a single technology (e.g., P4 with 2STiC) and are typically only accessible by some researchers.

**We call for a structural investment in state-of-the-art CompSys infrastructure**, both computer systems and networking devices, **that is co-developed and co-used in a triple-helix context, that is, in a collaboration by academia, government, and industry**. This infrastructure can be considered a fieldlab, like for example the Do IoT fieldlab[14], where the goal is to boost innovation and where society can try out their ideas. The distinctive difference with many other fieldlabs, is that this infrastructure would not focus on one technology, like 5G or GPUs, but computer systems and networks in general (as we argue in this manifesto comprises many disciplines and calls for a holistic design approach), with an emphasis on it becoming a trustworthy, responsible, and sustainable infrastructure.

By purpose, **this infrastructure has to operate differently from a normal production infrastructure**. This infrastructure will have a stable environment used for the various demonstrators, use cases, proofs of concept, and pre-competitive trials, and thus support technology and business developments at various levels of technological readiness (esp. working for TRLs 1-7); this matches the normal production purposes, albeit with a strong focus on upcoming, experimental hardware and software. But the infrastructure will also operate as a "sand-box" environment, where new research ideas can be tried without affecting the users, and where researchers can "break" the environment and explore limit-cases without fear of endangering "production." This sand-box mode is extremely valuable: CompSys is one of the few remaining sciences where grand experiments are envisioned and can actually be tried in practice; similarly, tried solutions that work can be merged into the stable environment, and thus push the state-of-the-art quickly over time. And fieldlab infrastructure can provide access to a realistic environment for the purposes of education and life-long learning.

Pragmatically, we recommend that building an experimental infrastructure for CompSys in the Netherlands should **not be a one-time investment, but rather a long-term project comprising tens to hundreds of millions of Euros**. Existing links with EOSC, GEANT, ESFRI/SLICES, and Gaia-X, and links of similar nature in the future, could greatly extend the capabilities and impact of this infrastructure, and could *democratize this science*.

**Without this investment, the Netherlands risks being excluded from the emerging CompSys research infrastructure funded by the EU**. Building and attracting our own expertise is a prerequisite for programs such as EuroHPC; future joint-undertakings require that matching-funds for research infrastructure are provided by each participating nation.

---

[14] https://doiotfieldlab.tudelftcampus.nl/

## R4. Recognize and fund cross-disciplinary translational CompSys research, and open ICT science.

The emergence of translational research in computer science [30] promises to change how the entire field operates. Early adopters could reap enormous benefits, much in the same way in which early innovators in translational medicine have changed the economic and societal structure of the field.

We recommend that **cross-disciplinary translational CompSys research should receive immediate attention and funding, commensurate with the high reward potential**.

Relatively to the current organizational and funding structures of CompSys in the Netherlands, cross-disciplinary translational research in CompSys is disruptive. Large-scale projects, e.g., aligned with the Nationale Wetenschapsagenda, could focus on various types of translational research and thus not have to include applied research in their program; correspondingly, applied research projects could link to other forms of translational research, rather than having to artificially justify their link to fundamental research. Cross-disciplinary research, focusing on CompSys technology or processes, should be funded; so should replication studies, which are common in some areas of science but exceedingly rare in computer science and in particular in CompSys science. The areas we have identified in Section 2, '[Cross-Disciplinary Translational CompSys Research](#)', could receive necessary attention and create unique opportunities for science and technology.

Concrete changes related to translational research include the addition of software engineers (scientific programmers) into the funding schemes, related to the type of research project and proportional to their size, but separately from the regular budget-limits currently considered by NWO programs.

Through the extended discussion on academic recognition (Erkennen en Waarderen), there is already room for more diverse research careers in the Dutch academia. However, we recommend making explicit that publishing artifacts resulting from translational research, e.g., FAIR and open software and data, and helping with their adoption, should be further recognized by both academic institutions and in NWO funding schemes. This will improve the career development prospects especially for young researchers.

## R5. Harmonize the national curriculum on CompSys topics.

Linked with the next recommendation, we recommend the **harmonization (unification, to the extent desirable and possible, retaining both common and unique aspects) of the national curriculum on CompSys topics**. This is a pragmatic recommendation, motivated by an increasing number of students, and by the high workload and expensive infrastructure associated with each student in this field relative to more theoretical subjects in computer science. The main benefit of a harmonized curriculum is that young professionals will learn the same common principles and vocabulary.

A curriculum harmonization would also allow us to teach the basic material more systematically, sharing expertise and resources, and would free up resources for advanced and especially unique learning objects. Teachers from *university X* could help with the influx of students at *university Y*. Cohorts and classrooms of reasonable size could thus be achieved for at least the more challenging aspects of the curriculum, which require much interaction, apprenticeship, and lab work for students to understand deeply. (A similar model exists for other core subjects in higher education, e.g., mathematics.)

If the recommendation to extend CompSys education and training to all levels of our society, and in particular to early age (that is, school and high-school), is implemented, then having a unified learning trajectory for this age, focusing on *computer systems literacy*, would be desirable as well.

## R6. Extend CompSys education and training to all levels of our society.

The digitalization of the Dutch society is increasing, and we foresee it to be irreversible. The magnitude of the change can be exemplified through a simple contrast between the level and type of societal infrastructure and services used a decade ago, and what we use today. Currently, computer intuition and often also knowledge have become necessary not only to run the economy, but also for daily activities such as doing bank transfers, laundry, and taxes, logging into government portals to make use of social services, using public transportation, making doctors' appointments, and talking to family and friends. Hence, it is of utmost importance to (**1**) include everyone in the educational process early, and (**2**) ensure life-long learning and training opportunities for all.

Although the use of digital goods and services is widespread in the Netherlands [36], exposure to computer science, let alone the CompSys area, is often delayed to the ternary and other higher-education institutes. There are very limited opportunities for high-school students to learn and educate themselves about computer science principles, approaches, processes, and applications. In contrast, computer science has emerged as a core subject in the basic education curriculums in countries like India, China, USA, and even European countries like the UK, Estonia, and Romania. Such initial exposure is important to develop a society-wide "CompSys temperament" similar to the "scientific temperament" that we expect from the Dutch population. A school-level computing and networking curriculum can include exposure to problem solving, programming, skills building, experiments and evaluation of ICT infrastructure, and computing processes and services. For example, Simon Peyton-Jones, a pioneer of computing, has led the UK-wide push to include computer science in primary and secondary education for all students and establish Computer Science as a Secondary Education academic qualification (GCSE) [38]. From their manifesto [38]: "**Computer Science is a foundational subject discipline, like maths and natural science, that every child should learn from primary school onward**".

We recommend to **ensure CompSys education and training become accessible to all levels of our society, for all**. Points of focus include: (**1**) Include computer science education in primary and secondary education in a nationwide systematic policy push; (**2**) Develop a computer science study curriculum for schools (ages 6-16), as core courses or as alternatives to other early STEM education; (**3**) Encourage diverse people to participate in computing; (**4**) Offer opportunities for further education and training for life-long learning in computer systems; (**5**) Because CompSys is an experimental science, this also implies investing in the required infrastructure for laboratories; and (**6**) Similarly to other design-first areas, foster a culture of making and breaking, through maker-spaces.

The need for an inclusive digitalization of the society is already recognized at the level of Government and Secretaries of State, and increasingly also by ministries (e.g., MinBZK, MinEZK). An explicit push to make CompSys diverse and open is required (see also R7), because the Netherlands cannot afford to leave knowledge, access, and exposure to computing and networking as just a hobby or only to a select few. Such push at an early stage of education is also critical to make computing and networking open and inclusive for all. In such a system, a diverse set of students (for example, by gender, socioeconomic background, and immigration background) can be helped to discover and develop CompSys skills together. Initiatives in this direction have appeared in the Netherlands, from the IPN EDI working groups (Alice and Eve exhibition, Celebrating women in computing), the VHTO school outreach, the Hedy Programming language in schools (Felienne Hermans), the inclusive Honours Programme at the VU, but a systematic nationwide policy and directions are still missing.

We advocate for **life-long learning on CompSys topics.** Computing and networking ideas and concepts are being developed at a very fast pace and look very different when checking at intervals of only 5-10 years. The rate of innovation is one of the primary reasons for the success of computer systems in our daily lives, but we must also ensure that the people who enter the Dutch economy are not left behind as the CompSys field moves forward. Offering regular training sessions (public-private partnership), opportunities for further education, public lectures, workloads, popular culture books, and summer and bootcamps is becoming necessary.

### R7. Link investment in research to hiring from under-represented groups, to foster diversity.

The authors and co-signatories of this document are firmly committed to improve the diversity of this field. One clear step in this direction is that investment in research should be tied to clear targets for hiring diverse people, e.g., from under-represented groups.

Responding to the requests for investment made in recommendations R1, R2, and R3, which are all sizable, would enable the Dutch Government to provide leadership in fostering diversity. Funding through the *Groeifondsprogramma*, in particular, could be tied to including diversity goals. The resulting network and people hired through this program would provide a significant step forward for diversity in CompSys in the Netherlands.

### R8. Ensure alignment with and integration into EU technology platforms and societal goals.

Geopolitically, the Netherlands is tightly integrated in the European projects proposed by the EU. However, the digital world offers new opportunities, new frontiers, and new political choices. Instead of aligning itself with emerging European technology platforms, the Netherlands might align with the US Big Tech companies or choose to rely on technology developed in China.

Our recommendation is direct: **the Netherlands should focus on the long-term integration into EU technology platforms and on full alignment with the EU societal goals**. Although these alignments will be initially costly, for the time-horizon 2035 it will deliver digital sovereignty from technology platforms that are not necessarily aligned, among others, with the economic interests of the EU, with taxation in the Netherlands on par with other value-adding industries, with respect for human rights, and with engagement against climate change. In turn, this form of digital sovereignty will help uphold the values and norms that define the Dutch society, including social inclusivity, equality, and fairness.

The cost of integration and alignment is that the Netherlands must co-sponsor EU activities such as those described in Section I.2.4, 'Digitalization in the Netherlands and aligned with the EU'. Investing in the European Green Deal, in EuroHPC and the 5G Toolbox, in the EOSC and Gaia-X platforms, may not come cheap initially and will require continuous funding. However, the resulting economic and societal benefits, including *brain mobility* to the Netherlands (which is historically a destination of such mobility), would more than compensate for the cost – see the box 'Economic and societal impact of CompSys in the Netherlands'.

## R9. NWO should introduce an ICT (infrastructure) plan in every NWO proposal that relies for its success on ICT infrastructure, matching the existing requirement for a data plan, to be evaluated by experts.

The NWO mandates data management plans in all competitive programs, aimed to ensure used and generated data is freely (FAIRly) accessible, thus aiming to make research more reproducible. The introduction of the data management plan has improved the process of publishing scientific data and contributes to open science in general. The data plan is organized as follows. Costs related to data management are eligible for funding. The data management plan must be developed in consultation with an expert (a "data steward"), typically located at the same institution as the applicant. Although the data management plan is not considered in the decision to fund the application, the application requires such a plan, the reviewers can give feedback that the project officer considers, and the project cannot start without the data management plan being approved by NWO.

Currently, *no similar plan exists for ICT infrastructure and services*, presumably because they are often considered a "solved issue". This happens even when the project relies on large hardware and software resources, or on infrastructure and services that are under active research in the community and/or are not commercially available. This is particularly surprising given that (1) the Netherlands has two active organizations whose core mission is to develop, implement, and facilitate access to ICT infrastructure and services for researchers, namely SURF and the Netherlands e-Science Center, and (2) each academic institution has a team focusing on ICT infrastructure and services.

We recommend that **NWO should require for *all* the grants, in the long term and after a gentle transition, a plan for ICT infrastructure and services** (i.e., ICT plan). The ICT plan should follow a similar approach as the data management plan. And, like for the data management plan, the ICT plan can be proportional in length to the reliance of the project on ICT infrastructure (and thus void if no such infrastructure is required). Thus: (**1**) the plan should be developed in consultation with an expert, (**2**) reviewers can give feedback that the project officer will consider, and (**3**) the project cannot start until this plan is approved by NWO. Unlike the data management plan, where the main concern is long-term data preservation *after* the research has occurred, research depending on ICT infrastructure and services *cannot occur at all without it*. Thus, we recommend that the mandate further specifies that (**4**) the plan on ICT infrastructure and services is actively considered for funding decisions, including refusing to fund proposals where the planned ICT resources are insufficient and/or infeasible to acquire.

**To facilitate adoption**, the ICT plan should be structured and streamlined in the NWO processes, such that it is not considerably longer or more difficult to fill in than the data management plan. **To transition** to this mandatory status, we recommend that the ICT plan is initially an optional element in new proposals, while being actively promoted and strongly encouraged *for proposals above a 100KEUR funding budget.* The transition could end within 5 years.

Finally, **to make the ICT plan feasible and efficient** for proposals where the ICT plan is required, we recommend allowing proponents to consider the ICT plan costs as a separate funding element, and, for efficiency and sustainability reasons, to actively incentivise and support proponents to use already available ICT infrastructure and services, through organizations like SURF and the Netherlands e-Science Center.

## R10. Align strategic investments in computing and networking in the Netherlands.

Various technology areas considered essential in the Dutch economy and society for the future decades, like AI, GovTech, and quantum, are experiencing rapid advances and increased competition. And for good reason, as these technologies are clearly promising. Yet, not all aspects of these technologies receive attention and, in particular, the need for advanced CompSys infrastructure is often overlooked or simply assumed available. We see here an opportunity for the Netherlands, to distinguish itself as a key player in AI and quantum that looks at the entire chain. After all, AI cannot run without computer systems, GovTech relies on the ability to collect, process, and share citizen data, and even the quantum internet relies on non-quantum computer systems and networks.

We recommend making strategic investments to **form strong links between CompSys and AI, between CompSys and Quantum, and between CompSys and GovTech**, such that the Netherlands can position itself as a key player who has digital sovereignty over all (critical) aspects of digitalization instead of only a few building blocks. One of the ways to achieve this is to pursue R2 and establish a CompSys coalition, which could then serve as liaison with the similar coalitions, e.g., in AI and quantum. Another is to make sure large grants from the Government include clauses that incentivize each of these directions to link to the others.

To prevent that progress becomes too slow due to delays in achieving strategic and executive alignment, different fields could develop independently but with a coordinating, cross-disciplinary alliance that would aim to reconcile the differences and forge alignment in the mid-term (by 2035).

## Conclusion

Computer systems and networks (CompSys) are the backbone of the digital economy. This CompSys Manifesto has explained the need for CompSys research and education. This enables the moonshot of transitioning to a fully digitalized Dutch society: the grand societal challenge of digitalizing the country and linking it to the European ICT platforms, while tailoring to what the society expects from ICT infrastructure, namely maintainability, responsibility, and sustainability.

We have proposed a groundbreaking, holistic, and collective technology roadmap, where the high quality of CompSys researchers in the Netherlands can leverage and extend their global content-related leadership. We have explained that Dutch CompSys research could continue to deliver remarkable and outsized contributions to the Dutch society, if linked with translational CompSys research, where cross-disciplinary uses of CompSys technology can create new, groundbreaking opportunities for a variety of sciences and domains, including the growing fields of AI, machine learning, and quantum computing. We have also identified key opportunities for Dutch higher education, which should truly service all in our society.

Looking at horizon 2035, to maximize impact, CompSys research in the Netherlands requires the formation of a national CompSys coalition, and immediate and sustained investment. The field can only hope to deliver, and the fields that rely on it, such as computational sciences, will only continue to prosper, if cross-disciplinary translational CompSys research is immediately placed on the government's agenda for investment. The field also requires systemic investments in CompSys infrastructure for education, research, demonstrations, and proof-of-concepts.

To promote diversity in this critical field for the Dutch society, we also recommend linking investment in CompSys research to hiring from under-represented groups – in particular, women, minorities, and recent immigrants –, and at stimulating diversity wherever the government provides funding.

**The 10 recommendations, in summary:**

1. Invest in personnel for holistic CompSys research: a structural investment of €195 million over the next 10 years, co-financed by the Government and by industry stakeholders.

2. Establish and maintain a coalition for CompSys, through a direct investment of €15 million over the next 10 years.

3. Invest in CompSys infrastructure for research, demonstrations, proof-of-concepts, and pilots: a structural investment of €145 million over the next 10 years.

4. Recognize and fund cross-disciplinary translational CompSys research and open ICT science.

5. Harmonize the national curriculum on CompSys topics, to the extent possible.

6. Extend CompSys education and training to all levels of our society. Be inclusive.

7. Link investment in research to hiring from under-represented groups, to foster diversity.

8. Ensure alignment with and integration into EU technology platforms and societal goals.

9. Introduce an ICT plan in NWO proposals that rely for their success on ICT infrastructure, with form following that of the data-management plan.

10. Align strategic investments in computing and networking in the Netherlands.

## Signatories

This manifesto represents the **combined views** of tens of specialists, clients, and societal stakeholders, raising awareness about challenges and opportunities in computer systems and networking (CompSys) research in the Netherlands toward shaping a national strategy in collaboration with the government, funding agencies, and other stakeholders. Although this manifesto is the first combined view in the Netherlands on computer systems and networking research and education in more than two decades, the experts that co-sign this manifesto have recently published many peer- and expert-reviewed papers on the key topics, thus obtaining international validation and recognition for the views expressed in this document.

We remain open to extending the support base for this manifesto. To work together on the themes of this manifesto, we welcome you to the IPN SIG on Future Computer and Network Systems (FCNS). If you are interested in this initiative, please contact us.

**Editors**: Alexandru Iosup (Vrije Universiteit Amsterdam, VU, A.Iosup@vu.nl) and Fernando Kuipers (Delft University of Technology, TU Delft, F.A.Kuipers@tudelft.nl).

**Primary contributors**, in alphabetical order:
Paola Grosso (Universiteit van Amsterdam, UvA), Alexandru Iosup (VU), Fernando Kuipers (TU Delft),
Francesco Regazzoni (UvA), Jan Rellermeyer (TU Delft), Animesh Trivedi (VU),
Alexandru Uta (U. Leiden), Ana Lucia Varbanescu (UTwente and UvA), and Lin Wang (VU).

**Co-signatories** from the Netherlands, in alphabetical order:
Roel Aaij (NIKHEF), Benny Akesson (UvA/TNO), Henri Bal (VU),
Suzan Bayhan (U. Twente), Vincent van Beek (Solvinity), Adam Belloum (UvA),
Peter Boncz (CWI/VU), Pablo Cesar (CWI/TU Delft), Valeriu Codreanu (SURF),
Henk Corporaal (TU/e), Jérémie Decouchant (TU Delft), Yuri Demchenko (UvA),
Marten van Dijk (CWI), Aaron Ding (TU Delft), Sagar Dolas (SURF),
Dick Epema (TU Delft), Georgi Gaydadjiev (RUG), Clemens Grelck (UvA),
David Groep (NIKHEF), Said Hamdioui (TU Delft), Paul Havinga (U. Twente),
Sonia Heemstra (TU/e), Geert Heijenk (U. Twente), Cristian Hesselman (SIDN Labs),
Georgios Iosifidis (TU Delft), Cees de Laat (UvA), Guohao Lan (TU Delft),
Zoltan Mann (UvA), Johan Lukkien (TU/e), Omar Niamut (TNO),
Rob van Nieuwpoort (NLeSC/UvA), Pieter Nooren (TNO), Chrysa Papagianni (UvA),
Anuj Pathania (UvA), Przemysław Pawełczak (TU Delft), Andreas Peter (U. Twente),
Andy Pimentel (UvA), Aske Plaat (U Leiden), Damian Podăreanu (SURF),
Ronald van der Pol (SURF), Aiko Pras (U. Twente), Kristian Rietveld (U. Leiden),
John Romein (ASTRON), Georgios Smaragdakis (TU Delft), Andries Stam (Almende),
Maarten van Steen (U. Twente), Todor Stefanov (U. Leiden), Sander Stuijk (TU/e),
Ranga Rao Venkatesha Prasad (TU Delft), Peter Verkoulen (Gaia-X NL), Qing Wang (TU Delft),
Paul Wijngaard (TNO), Anton Wijs (TU/e), Zhiming Zhao (UvA), and
Marco Zuniga (TU Delft).

## Who Is Who in CompSys in the Netherlands?

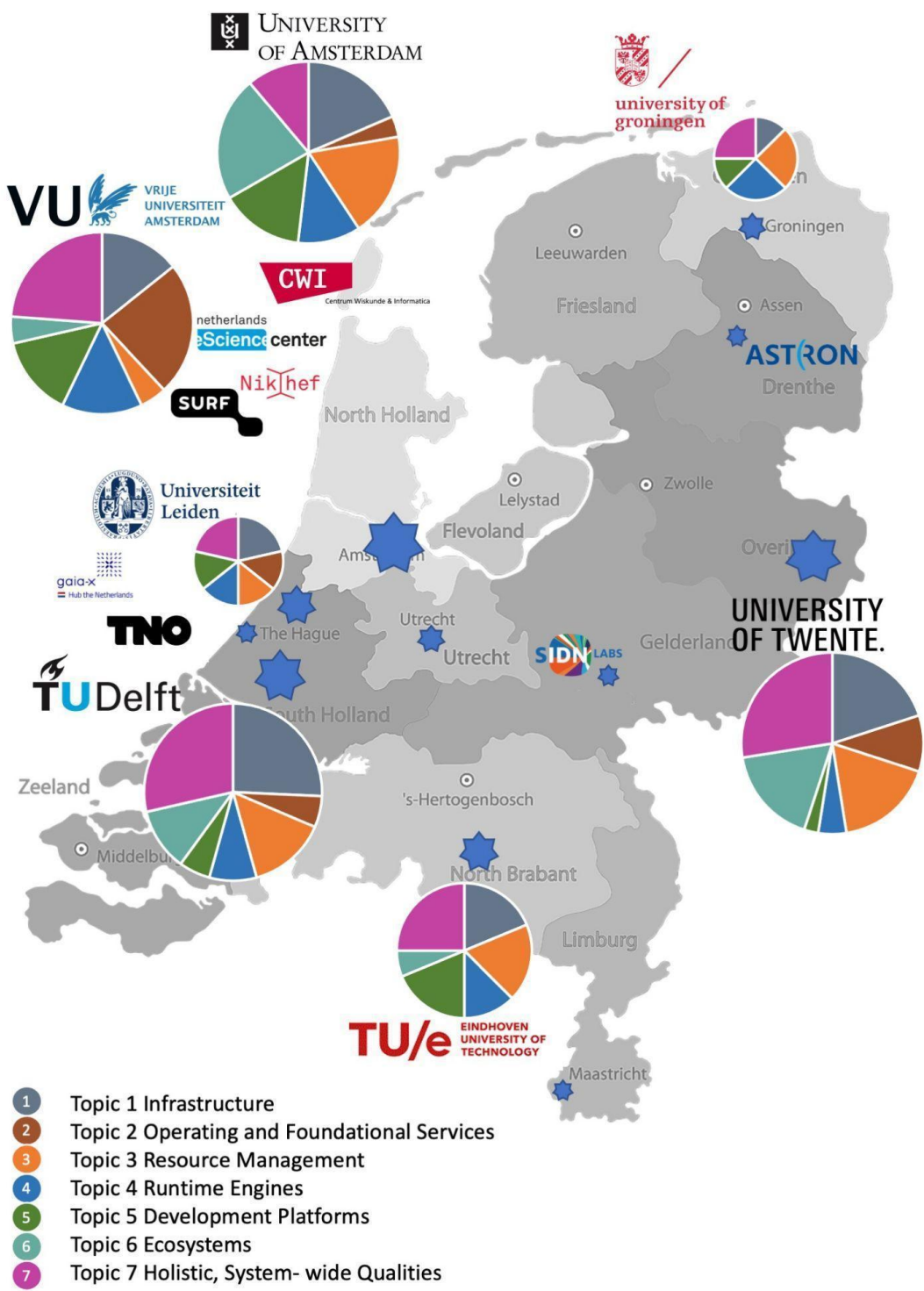

More details about who's who: https://bit.ly/CompSysNLWhosWho